\documentclass[preprint,10pt]{article}

\usepackage[margin=1.0in]{geometry}

\usepackage[utf8]{inputenc}
\usepackage{amsmath}
\usepackage{amssymb}
\usepackage{graphicx}
\usepackage{url}
\usepackage{wasysym}
\usepackage{upgreek}
\usepackage{longtable}
\usepackage{breqn}
\usepackage{stmaryrd}
\usepackage{pdfpages}
\usepackage{booktabs}
\usepackage{tabularx}
\usepackage{lipsum}
\usepackage{breqn}
\usepackage{url}

\usepackage{breakurl}
\usepackage[breaklinks]{hyperref}

\usepackage{amsmath}
\usepackage{amssymb}
\usepackage{wasysym}
\usepackage{amsthm}
\usepackage{tikz}
\usepackage{longtable}
\usepackage{algorithm}
\usepackage{wasysym}
\usepackage[noend]{algpseudocode}
\usepackage{tipa}
\usepackage{amsfonts}
\usepackage{calligra}
\usepackage{calrsfs}
\usepackage{lineno}
\usepackage{marvosym}
\usepackage{subcaption}
\usepackage{booktabs}
\usepackage{multirow}
\usepackage{bm}

\hypersetup{
    colorlinks = true,
    linkbordercolor = {white},
    linkcolor = {red},
	urlcolor = {magenta},
	citecolor = {red},
}

\def\vector#1{\underline{\text{#1}}}
\def\tso#1{\underline{\underline{\text{#1}}}}
\def\tfo#1{\underline{\underline{\underline{\underline{\text{#1}}}}}}
\def\singlecontraction{\cdot}
\def\doublecontraction{\,:\,}
\def\grad{\underline \triangledown\,}

\def\pu{\underline{\lambda}_u}
\def\punu{\underline{\lambda}_{u_{\nu}}}

\def\Bab{B_{\alpha \beta}}
\def\bab{b_{\alpha \beta}}

\def\domega#1{\partial \Omega_{#1}}

\def\Gone{\vector{G}_{1}}
\def\Gtwo{\vector{G}_{2}}
\def\Gthree{\vector{G}_{3}}
\def\Galpha{\vector{G}_{\alpha}}

\def\gone{\vector{g}_{1}}
\def\gtwo{\vector{g}_{2}}
\def\gthree{\vector{g}_{3}}
\def\galpha{\vector{g}_{\alpha}}

\def\Aone{\vector{A}_{1}}
\def\Atwo{\vector{A}_{2}}
\def\Athree{\vector{A}_{3}}
\def\Aalpha{\vector{A}_{\alpha}}

\def\aone{\vector{a}_{1}}
\def\atwo{\vector{a}_{2}}
\def\athree{\vector{a}_{3}}
\def\aalpha{\vector{a}_{\alpha}}

\newcommand{\ddt}[2]{\frac{\partial #1}{\partial #2}}

\makeatletter
\newenvironment{breakablealgorithm}
  {
   \begin{center}
     \refstepcounter{algorithm}
     \hrule height.8pt depth0pt \kern2pt
     \renewcommand{\caption}[2][\relax]{
       {\raggedright\textbf{\ALG@name~\thealgorithm} ##2\par}%
       \ifx\relax##1\relax 
         \addcontentsline{loa}{algorithm}{\protect\numberline{\thealgorithm}##2}%
       \else 
         \addcontentsline{loa}{algorithm}{\protect\numberline{\thealgorithm}##1}%
       \fi
       \kern2pt\hrule\kern2pt
     }
  }{
     \kern2pt\hrule\relax
   \end{center}
  }
\makeatother

\begin{document}

\title{Isogeometric Analysis of Elastic Sheets Exhibiting Combined Bending and Stretching using Dynamic Relaxation}
\author{Nikhil Padhye and Subodh Kalia}

\date{}
\maketitle

\abstract

Shells are ubiquitous thin structures that can undergo large nonlinear elastic deformations while exhibiting 
combined modes of bending and stretching, and have profound modern applications. In this paper, we have proposed a new
Isogeometric formulation, based on classical Koiter nonlinear shell theory, to study 
instability problems like wrinkling and buckling in \textit{thin} shells. 
The use of NURBS-basis provides rotation-free, conforming, higher-order spatial continuity, such that
 curvatures and membrane strains can be 
computed directly from the interpolation of the position vectors of the control points. 
A pseudo, dissipative and discrete, dynamical 
system is constructed, and static equilibrium solutions are obtained by the method
of dynamic relaxation (DR). 
A high-performance computing-based implementation of the DR is presented,
and the proposed formulation is benchmarked against 
several existing numerical, and experimental results. The advantages of this formulation, over traditional finite element 
approaches, in assessing structural response of the shells are presented. \\\\


\section{Introduction}
\begin{figure}[ht]
    \centering
    \includegraphics[scale=0.4]{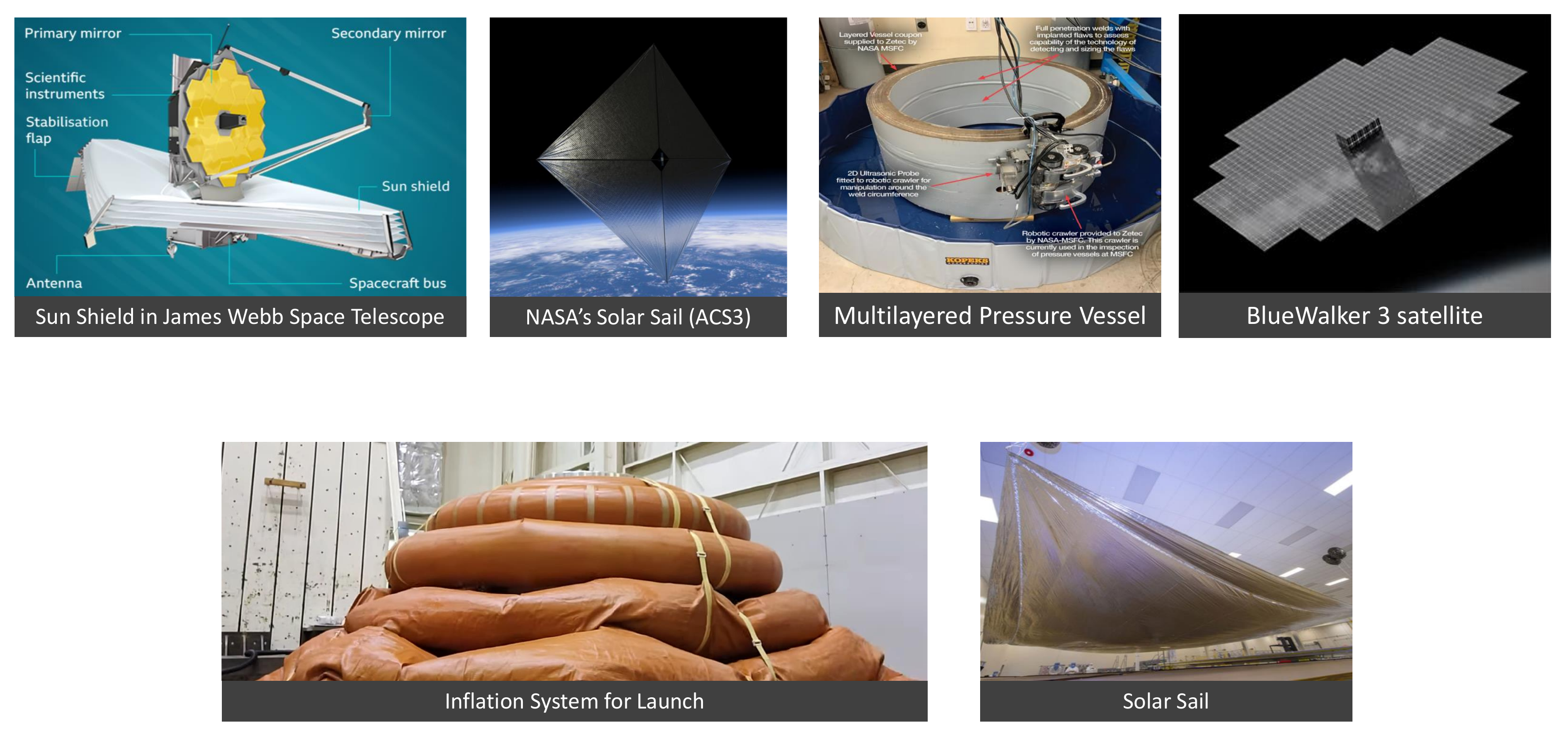}
    \caption{Modern space-related applications of shells, plates, and membranes \cite{nasa-1,nasa-2,nasa-3,nasa-4,nasa-5,nasa-6}.}
    \label{nasa}
\end{figure}

A shell is a three-dimensional structure with thickness much smaller compared to 
the other characteristic dimensions. Modeling the mechanics and deformation behavior 
of shells has attracted extensive research since the middle of the 19$^{th}$ century due to their broad
usefulness in structural engineering. 
Shells with a flat midsurface are known as plates and, membranes represent the limiting case of 
extremely thin shells. A good shell model can be used to model a plate or membrane behavior, but not vice-versa. 
Shells are well-known to provide light-weight and stiff load-bearing capacities that enable 
design and functioning of complex engineering structures, and facilitate actuation. 
Widely encountered examples of shells (and plates, or membranes) include (but are not limited to):  
body parts and components in airplanes, ships, automobiles, and space ships, armors, bridges and buildings, solar sails, lipid membranes,
biological cells, crystal films, fabrics, stretchable electronics, soft robotics,  
oragami applications, plastic bags and packaging materials, nano-devices, two-dimensional or 
meta-materials, plants, living organisms, and human organs, 
\cite{bartels2017bilayer, bonito2021numerical,ghaffari2018new,bartels2018modeling,turco2016fiber,boncheva2005magnetic,chung2018reprogrammable,dias2012geometric,johnson2011status,blachut2014experimental,
hilburger2012developing,caspar1962physical,dresselhaus1998physical,bala2005non}, and space applications (Figure ~\ref{nasa}).
Given the ubiquitous presence of shells, reliable computational models are necessary to predict the deformation behaviour so that shells with optimal functionalities 
can be designed, and equally (if not more importantly) the mechanical response of shells in natural settings can be understood to enable nature-inspired design. 
E.g., light-weighting and stiffening properties of a shell are conflicting in nature, because decreasing the shell thickness will reduce its weight but also increase its compliance, 
therefore for designing an optimal structure, with low weight and high load-bearing capacity, a reliable predictive model is necessary during the design phase.  
The subject matter considered in this work is \textit{nonlinear} elastic behavior of materially uniform shells without any pre-strains 
(i.e. shells with stress free reference configuration). Plastic deformation, damage, and fracture are also not discussed. 

The goal of any shell theory (or a 2D model) is to capture the deformation of a thin structure in terms of its midsurface 
deformation, while making appropriate through-thickness approximations. The motivation for this
undertaking arises from the fact that performing complete three-dimensional (nonlinear) deformation analysis, by treating a thin structure
as 3D continua, in general, is intractable. Furthermore, in context to computational finite elements,
the 3D finite element analysis of a large thin structure can be quite expensive, whereas 2D finite element shell models
can be evaluated relatively efficiently. The key question always worth asking in context to a given shell model, and associated finite element (FE) discretization
is: What is the range of applicability of the given shell model (and the associated finite element discretization) for different material behaviors, shell geometries, and loading scenarios?
From a theoretical standpoint, one must also ask for a justification of the given shell model \textit{vis-{\'a}-vis} three-dimensional nonlinear elasticity, i.e., 
what is the connection, and accuracy of a particular shell model with respect to complete three-dimensional nonlinear elastic treatment. Additional consideration, for practical purposes, 
is the complexity or computational cost of the given shell model (and its FE discretization).  
There are numerous nonlinear shell models in the literature, notwithstanding their representative success on the chosen sample problems, a large number of these models 
do not have any kind of rigorous justification, rather they incorporate  \textit{ad hoc} assumptions in the model formulation, and its FE discretization.  
In case of linear elastic shells (the case of small elastic deformations), these aforementioned issues have been addressed more effectively; however, 
a general purpose rigorous nonlinear elastic shell modeling (and its finite element analyses) remains a challenging topic of research to date. 
An excellent account of these issues is given in \cite{ciarlet2022mathematical,ciarlet2000theory,ciarlet2008introduction}, and references cited therein. 

For \textit{thin} shells, a nonlinear model (with its variants) originally proposed by Koiter \cite{koiter1966nonlinear,koiter1960consistent}
has given satisfactory performance in various practical settings over last several decades. This nonlinear
shell model proposed by Koiter involves two \textit{a priori} assumptions: (1) Kirchhoff-Love kinematics: any material 
point located on a normal to the midsurface, in the reference configuration, 
will remain on the normal to the midsurface in the deformed configuration, and  
the distance of any such point from the midsurface remains constant after the deformation has taken place. 
(2) The state of stress in the shell is planar, i.e. stresses during deformation are parallel to the midsurface. The material behavior in Koiter's modeling is chosen as 
Saint Venant-Kirchhoff (SVK) material. Although the original Koiter model 
incorporated ad hoc kinematic assumptions, subsequently, rigorous justifications for the Koiter's model have been provided
by Ciarlet and co-workers, see in particular \cite{ciarlet2018nonlinear}; where the authors have proposed a relatively general nonlinear shell model 
(that can be used for any type of shell, without any restriction on the geometry of the surface or on the boundary conditions), 
for which existence of a solution under arbitrary loading is proven, and the proposed nonlinear shell model is shown to be ``formally asymptotically equivalent'' to the classical nonlinear Koiter model. See the general body of work 
in \cite{ciarlet2018nonlinear,mardare2019nonlinear,ciarlet2018existence,ciarlet2001justification,ciarlet1996asymptotic}.
Koiter's model is considered to be one of the best ``all-around'' model capturing combined bending and stretching of 
thin shells \cite{ciarlet2018nonlinear}. As stated in \cite{ciarlet2018nonlinear}, there are other two-dimensional nonlinear shell models 
that have been rigorously justified in the shell theory literature by existence theorems, such as, the membrane-dominated model \cite{le1996membrane}, 
the flexural-dominated model \cite{ciarlet1998existence,friesecke2003derivation}, and the two-dimensional model of Koiter's type for spherical and ``almost spherical shells'' 
\cite{bunoiu2015existence}, \cite{ciarlet2016mathematical}; however, rigorous justification for these models is only applicable  under restrictive conditions. Finally, 
we note that although classical nonlinear Koiter shell does not have a general result for an existence of solution within the conventional framework of calculas of variations, however, 
that does not imply that energy minimizers fail to exist for specific boundary value problems. 
In light of these considerations, since the Koiter's nonlinear shell model is ``justified'' for shells with \textit{small} thicknesses, i.e., thin shells, 
and for studying localized or extended wrinkling or buckling phenomena a model that simultaneously accounts for 
bending and stretching, as done in Koiter's model, is required \cite{cerda2002wrinkling,cerda2003geometry,vaziri2008localized}, 
we have chosen the classical nonlinear Koiter shell model in this work. Our proposed high-performance computational framework is general enough, however, so that more advanced
models, such as those in \cite{ciarlet2018nonlinear}, can be easily implemented.

Once an appropriate nonlinear shell model is chosen, a suitable finite element discretization scheme, and a numerical procedure to solve equilibrium configurations are needed. Depending on the choice of underlying shell model, 
several types of finite element formulations, each with their own advantages, limitations, and challenges, 
have been proposed in the literature. These range from employing traditional formulation with rotational and translational degrees of freedom, recently proposed 
rotation free formulations, higher-order ($C^1$) spatially continuous basis functions (either ensuring tangent continuity at discrete locations, or over the entire element), discontinuous galerkin methods, enhanced strain or mixed 
methods, e.g., see 
\cite{hughes1981nonlinear,chapelle2010finite,stolarski2013nonlinear,dau2006c1,campello2003triangular,brank2002nonlinear,buchter1994three,talamini2016discontinuous,cirak2001fully} and cited works therein for a general overview of the past approaches. 
Given that the mechanical response of a thin structure, in general, is susceptible to imperfections \cite{budiansky1966dynamic}, discretization errors (or geometrical artifacts) introduced during the creation of 
a finite element model can lead to erroneous predictions in structural analyses (especially those involving instabilities); therefore, a spatial discretization techniques that can eliminate discretization errors are highly desirable. Secondly, ensuring $C^1$ continuity in finite element basis functions, 
such that a particular component of the interpolated midsurface displacements can lie in $H^2$ Sobolev space, is challenging and difficult. (The  need for $C^1$ continuity arises
due to second-order derivatives of one of the components of the midsurface displacements  in the energy functional, whose stationarity leading to the equilibrium solution is sought.)
On the other hand, use of rotational degrees of freedom, as done in the literature, to avoid direct requirement of $C^1$ continuity in displacements, can lead to the issues of  locking when the 
shell thickness decreases. 
To this end, in the recent years, researchers have successfully used the accurate spatial discretization, and higher-order interpolation properties provided by the Isogeometric Analysis (IGA)  to model nonlinear shell behaviors \cite{kiendl2010bending,duong2017new}.
Isogeometric Analysis (IGA) has been proposed as a paradigm to potentially bridge the gap between design and 
analysis \cite{Hughes-book,bazilevs2006isogeometric,hughes2005isogeometric},
and has been particularly useful for analyses of shells. IGAs use control points (CPs) (which may or may not lie on the material surface), and special basis functions which can made arbitrarily smooth. 
Non-Uniform Rational B-Splines (NURBS) are the most popular basis functions in the IGA technology. The ``finite element model'' is automatically decided once a solid body is modeled using NURBs. In this work,
we have adopted NURBS-based Isogeometric Analysis to discretize the classical nonlinear Koiter shell for studying instability problems in thin shells. The past attempts on developing IGA-based methods for shells have not focused on studying instability problems like wrinkling, or buckling, and it is to this end, in this contribution, we have proposed the method dynamic relaxation, with IGA, to robustly solve equilibrium configurations. A major advantage of our formulation is that we do not require
``imperfection seeding'' in the initial structure, which in effect increases the structure's  compliance to a particular deformation mode that is ``seeded'',
for triggering buckling or wrinkling. A nonlinear elastic structure can exhibit multiple stable equilibrium solutions, therefore the requirement of 
 artificially seeding a structure with an ad hoc imperfection (size and shape) will strongly effect its deformation behaviour, and resulting equilibrium solution, and is 
 undesirable. Additionally, the method of dynamic relaxation is motivated by the concept Lyapunov stability, and creates a pseudo dynamical system with dissipation, that can 
 converge to a stable equilibria once the structure's kinetic energy is damped out. The DR does not require construction of tangent stiffness matrix, which can become ill- posed during instability. The past attempts of studying instability problems using DR, \cite{Taylor_Bertoldi_2014,nakashino2020geometrically,
nakashino2020geometrically,verhelst2019modelling,alic2016form}, have either used a finite difference method for the special case of plate, or  reported unsatisfactory results on sample problems (discussed further in Section ~\ref{sec:numerical-results}). Our proposed methodology 
 successfully reproduces a wide-range of instability results from the literature, and is scalable.

The rest of the paper is organized as follows: In Section ~\ref{sec:nonlinear-koiter-shell} we review the classical nonlinear shell model by Koiter, and present the principle of 
virtual work. Section ~\ref{sec:penalty-formulation} discusses penalty formulations for imposing Dirchlet boundary conditions, and enforcement of $C^1$ continuity at the mating external edges. Section ~\ref{sec:nurbs} summarizes the IGA-based shell modeling adopted in this study, along with numerical computation of 
various quantities required for finite element calculations. 
Method of dynamic relaxation for obtaining equilibrium solutions, and guidelines for its parameter selections are 
discussed in Section ~\ref{sec:dynamic-relaxation}. 
Finally, simulation results of the proposed formulation are presented in Section ~\ref{sec:numerical-results}.
In what follows, we have adopted the notation that 
($\singlecontraction$), and ($\doublecontraction$) represent single and double contractions, respectively.
Vectors and tensors are denoted by single and double underscores, respectively. Fourth order tensors are represented by 
quadruple underscores. Latin indices take values 1, 2, or 3, whereas
Greek indices take values 1, or 2. Einstein's summation convention is followed for repeated indices,  $\otimes$, and $\times$ represent
tensor product, and vector cross product, respectively. Vector fields after finite element discretization are represented as
column vectors, e.g.,  $\left[ \vector{x} \right]{_{3 \times 1}}$ denotes a column vector corresponding to the vector field 
\vector{x}. Transpose of a column vector is indicated with the symbol $^\top$. Euclidean norm of a vector is 
denoted by $||\cdot||$.

\section{Nonlinear Koiter Shell}
\label{sec:nonlinear-koiter-shell}
Koiter's nonlinear shell model can be obtained from the three-dimensional nonlinear elasticity by employing 
Kirchoff-Love (KL) kinematic hypothesis, and assuming SVK material behavior.
As stated before, according to the KL hypothesis, material points along the normal in the reference configuration will stay along the normal 
during deformation, and their distance from the midsurface remains unchanged. Secondly, 
transverse stresses are neglected. The SVK material choice is one of the simplest isotropic material models, and in three-dimensional elasticity 
its validity is usually limited to small strains. SVK material model is characterized by a fourth order elasticity tensor, which constitutively connects 
stress with Green-Lagrange strain tensor. 
For the sake of completeness, we first provide a concise derivation of the nonlinear Koiter shell model, and important kinematic assumptions that 
can affect the performance of this model to the class of considered problems, and the resulting
principle of virtual work. These are then used in the finite element formulation. Additional notation is explained
during the course of this study.

\subsection{Shell kinematics}
\label{shell-kinematics-derivation}
The shell midsurface is parameterized by curvilinear coordinates, $\theta^\alpha$ (where $\alpha$ = 1, 2), and 
$\zeta$ is used to denote the through-thickness coordinate. Any material point $(\theta^1, \theta^2, \zeta)$ is located by 
the position vectors $\vector{X}(\theta^1, \theta^2, \zeta)$, and $\vector{x}(\theta^1, \theta^2, \zeta)$ in the reference, and deformed configurations, respectively. The partial derivatives with respect to the curvilinear coordinates are denoted as 
$\{ . \}_{,\alpha} = \partial \{ . \}/\partial \theta^\alpha$. The covariant basis vectors at any point (away from or on the midsurface) in the reference configuration are represented by $\vector{G}_i$, with
\begin{equation}
\Galpha = \ddt{\vector{X}(\theta^1, \theta^2, \zeta)}{\theta^\alpha}, \quad \Gthree = \frac{\Gone \times \Gtwo }{|\Gone \times \Gtwo|}.
\label{ref_Gs}
\end{equation}
Similarly, the covariant basis vectors at any point (away from or on the midsurface) in the deformed configuration are represented by $\vector{g}_i$, with
\begin{equation}
\galpha = \ddt{\vector{x}(\theta^1, \theta^2, \zeta)}{\theta^\alpha}, \quad \gthree = \frac{\gone \times \gtwo }{|\gone \times \gtwo|}.
\label{def_gs}
\end{equation}
The contravariant bases in the reference, and deformed configurations are represented by $\vector{G}^i$, and $\vector{g}^i$, respectively. The contravariant basis vectors satisfy the following relations
\begin{align}
\vector{G}^i \singlecontraction \vector{G}_j &= \delta^{i}_{j},\\
\vector{g}^i \singlecontraction \vector{g}_j &= \delta^{i}_{j},
\label{covariant_def}
\end{align}
where $\delta^{i}_{j}$ is the Kronecker-delta. 
The reference and deformed metrics are defined as follows
\begin{align}
G_{ij}  &=\vector{G}_i \singlecontraction \vector{G}_j,\label{metric_def1}\\
g_{ij}	&= \vector{g}_i \singlecontraction \vector{g}_j.
\label{metric_def2}
\end{align}

The three-dimensional identity tensor ($\tso{I}$), and the deformation gradient ($\tso{F}$) are defined in terms of 
base vectors in the reference and deformed configurations as
\begin{equation}
\tso{I} = G_{ij} \vector{G}^i \otimes \vector{G}^j,
\label{3d_i}
\end{equation}
and
\begin{equation}
\tso{F} = \vector{g}_i \otimes \vector{G}^j.
\label{deformation_gradient}
\end{equation}
The three-dimensional Green-Lagrange strain ($\tso{E}$) is given as
\begin{equation}
\tso{E} = \frac{1}{2} \left( \tso{F}^\top \singlecontraction \tso{F} - \tso{I} \right),
\label{gl_strain}
\end{equation}
and using equations \ref{3d_i}, \ref{deformation_gradient} and \ref{gl_strain},
\tso{E} can be written as 
\begin{align}
\tso{E} &= \frac{1}{2} \left( g_{ij} - G_{ij} \right) \vector{G}^i \otimes \vector{G}^j \nonumber \\
&= E_{ij} \, \vector{G}^i \otimes \vector{G}^j,
\label{three_d_gl_strain}
\end{align}
with,
\begin{equation}
E_{ij} = \frac{1}{2} \left( g_{ij} - G_{ij} \right).
\label{E_ij}
\end{equation}
Since the Green-Lagrange strain tensor expressed in equation \ref{three_d_gl_strain} is based on 
three-dimensional continuum kinematics, we will now employ Koiter's kinematic assumptions across the shell thickness, 
to obtain specialized form of \tso{E}. 

In the reference configuration, the position vector of any point $\vector{X}(\theta^1, \theta^2, \zeta)$ (away from the 
midsurface) can be written in terms of position vector of a point on the midsurface $\vector{X}(\theta^1, \theta^2)$, and 
its distance $\zeta$ from the midsurface along the unit normal (denoted by $\Athree$) as
\begin{equation}
\vector{X}(\theta^1, \theta^2, \zeta) = \vector{X}(\theta^1, \theta^2) + \zeta \Athree.
\label{kinematic_assumption_ref}
\end{equation}
Since a material point located at ($\theta^1$, $\theta^2$, $\zeta$) in the reference configuration stays at a distance $\zeta$ along the unit normal (denoted as $\athree$) in the deformed configuration, its position vector can be written as 
\begin{equation}
\vector{x}(\theta^1, \theta^2, \zeta) = \vector{x}(\theta^1, \theta^2) + \zeta \athree,
\label{kinematic_assumption_def}
\end{equation}
where $\vector{x}(\theta^1, \theta^2)$ is position vector of a point on the shell midsurface (where $\zeta=0$) in the deformed configuration. We represent the covariant basis vectors at the shell midsurface, in the reference, and deformed midsurfaces, as 
$\vector{A}_1$, $\vector{A}_2$, and $\vector{A}_3$, and $\vector{a}_1$, $\vector{a}_2$, and $\vector{a}_3$, respectively. Thus, using equations \ref{ref_Gs}, and \ref{kinematic_assumption_ref} we have
\begin{equation}
\Aalpha = \ddt{\vector{X}(\theta^1, \theta^2)}{\theta^\alpha}, \quad \Athree = \frac{\Aone \times \Atwo }{|\Aone \times \Atwo|},
\label{ref_As}
\end{equation}
and using equations \ref{def_gs}, and \ref{kinematic_assumption_def} we obtain
\begin{equation}
\aalpha = \ddt{\vector{x}(\theta^1, \theta^2)}{\theta^\alpha}, \quad \athree = \frac{\aone \times \atwo }{|\aone \times \atwo|}.
\label{def_as}
\end{equation}

Considering the four components $E_{11}$, $E_{22}$,  $E_{12}$ and $E_{21}$ of the Green-Lagrange tensor 
($E_{ij}$), denoted by $E_{\alpha\beta}$, and ignoring the transverse strains transverse strains, ($E_{13}$,  $E_{23}$ and, $E_{33}$), using Equation ~\ref{E_ij}, we can write 
\begin{equation}
E_{\alpha \beta} = \frac{1}{2} \left( g_{\alpha\beta} - G_{\alpha\beta} \right).
\label{E_alpha_beta}
\end{equation}
Using Equations ~\ref{ref_Gs} and ~\ref{kinematic_assumption_ref}, and 
~\ref{def_gs} and ~\ref{kinematic_assumption_def}, we can write 
the reference and deformed metric components as
\begin{equation}
G_{\alpha\beta} = \vector{A}_\alpha \singlecontraction \vector{A}_\beta + \zeta (\vector{A}_\alpha \singlecontraction \vector{A}_{3,\beta} + \vector{A}_\beta \singlecontraction \vector{A}_{3, \alpha}) + \zeta^2 (\vector{A}_{3, \alpha} \singlecontraction \vector{A}_{3, \beta}),
\label{eq:Reference-Curvature}
\end{equation}
and 
\begin{equation}
g_{\alpha\beta} = \vector{a}_\alpha \singlecontraction \vector{a}_\beta + \zeta (\vector{a}_\alpha \singlecontraction \vector{a}_{3,\beta} + \vector{a}_\beta \singlecontraction \vector{a}_{3, \alpha}) + \zeta^2 (\vector{a}_{3, \alpha} \singlecontraction \vector{a}_{3, \beta}).
\label{eq:Deformed-Curvature}
\end{equation}

From the second fundamental form (Gauss-Weingarten equations), the reference and deformed shell midsurfaces 
curvature components, $\Bab$ and $\bab$, respectively, are given as 
\begin{align}
\Bab &= \Athree \singlecontraction \vector{A}_{\alpha, \beta} \nonumber \\
     &= - \vector{A}_{3,\alpha} \singlecontraction \vector{A}_\beta,
\label{Bab_def}
\end{align}
and 
\begin{align}
\bab &= \athree \singlecontraction \vector{a}_{\alpha, \beta} \nonumber \\
     &= - \vector{a}_{3,\alpha} \singlecontraction \vector{a}_\beta.
\label{bab_def}
\end{align}
Using Equations ~\ref{Bab_def} and ~\ref{bab_def} in Equations 
~\ref{eq:Reference-Curvature} and ~\ref{eq:Deformed-Curvature}, respectively, and ignoring 
higher-order ($\zeta^2$) terms (assuming that the shell is \textit{thin}), we obtain 
\begin{equation}
E_{\alpha \beta} = \frac{1}{2} (a_{\alpha\beta} - A_{\alpha\beta} ) + \zeta (\Bab - \bab).
\label{Eab}
\end{equation}
We will return to the issues of ignoring $\zeta^2$ term. 
By defining  $K_{\alpha\beta}$ as the change in the curvature of the midsurface, 
and $\varepsilon_{\alpha\beta}$ as the membrane strain as follows
\begin{equation}
K_{\alpha\beta} = B_{\alpha\beta} - b_{\alpha\beta},
\label{Kab}
\end{equation}
and
\begin{equation}
\varepsilon_{\alpha\beta} = \frac{1}{2} (a_{\alpha\beta} - A_{\alpha\beta}),
\label{var_eab}
\end{equation}
we can rewrite Equation ~\ref{Eab} as 
\begin{equation}
E_{\alpha\beta} = \varepsilon_{\alpha\beta} + \zeta K_{\alpha\beta}.
\label{eab}
\end{equation}
Clearly, in the above equation, the mid-surface curvature introduces the through-thickness strain variation. We note that the strain components 
$E_{\alpha\beta}$, given in Equation ~\ref{eab}, are with respect to reference contravariant basis (see Equation
~\ref{three_d_gl_strain}).

\subsection{Constitutive Behavior}

The strain energy density ($W$) according to SVK material model, in terms of the Green-Lagrange strain tensor is given as
\begin{equation}
W(\tso{E}) = \frac{\lambda}{2} \Big[ \text{tr(} \tso{E} \text{)} \Big]^2 + \mu \,\, \text{tr(} \tso{E} \singlecontraction \tso{E} \text{)},
\label{SVK_model}
\end{equation}
where $\lambda$ and $\mu$ are Lame's parameters. In terms of Young's modulus ($E$) and Poisson's ratio ($\nu$) are given as
\begin{equation}
\lambda = \frac{E \nu}{(1+\nu)(1-2\nu)} \quad  \text{and} \quad \mu = \frac{E}{2(1+\nu)}.
\label{lamesparameters}
\end{equation}

The Second Piola-Kirchoff stress tensor ($\tso{S}$) being work-conjugate to $\tso{E}$ (Green-Lagrange strain), is then given as 
\begin{equation}
\tso{S} = \frac{\partial W}{\partial \tso{E}}.
\label{S_stress}
\end{equation}
Using Equations \ref{S_stress} and \ref{SVK_model}, the Second Piola-Kirchoff stress can be written as
\begin{equation}
\tso{S} = 2 \mu \, \tso{E} + \lambda \, \text{tr(} \tso{E} \text{)} \tso{I},
\label{S_stress_expr}
\end{equation}
and the Cauchy stress ($\sigma$), related to Second Piola-Kirchoff stress, as
\begin{equation}
\tso{$\sigma$} = \frac{1}{J} \left( \tso{F} \singlecontraction \tso{S} \singlecontraction \tso{F}^\top \right),
\label{cauchy_stress}
\end{equation}
where $J = \text{det(}\tso{F}\text{)}$. 
The constitutive relationship in Equation \ref{S_stress_expr} can also be written as
\begin{equation}
\tso{S} = \tfo{C} \doublecontraction \tso{E},
\label{s=ce}
\end{equation}
where $\tfo{C}$ is the fourth order elasticity tensor. 
Since the representation of the components of the elasticity tensor ($\tfo{C}$) is highly simplified in an 
orthonormal basis, we chose a local orthonormal basis $\vector{E}_1$, $\vector{E}_2$, and $\vector{E}_3$, and write 
\begin{equation}
\tfo{C} = C_{ijkl} \vector{E}_i \otimes \vector{E}_j \otimes \vector{E}_k \otimes \vector{E}_l,
\end{equation}
where 
\begin{equation}
C_{ijkl} = \lambda \delta_{ij} \delta_{kl} + \mu (\delta_{ik} \delta_{jl}  + \delta_{il} \delta_{jk}),
\end{equation}
$\delta_{ij}$ is the Kronecker delta. 
Since the components of \tfo{C} have been expressed in a local orthonormal basis, to be able to use Equation ~\ref{s=ce} 
component wise, we need to express the components of the Green-Lagrange tensor ($\tso{E}$) in this local orthonormal basis. 
We note that the reference contravariant basis vectors $\vector{G}^1$ and $\vector{G}^2$ can vary along the shell midsurface, and need not be orthogonal. We select  $\vector{E}_1$, and $\vector{E}_2$, mutually orthogonal in the plane formed by 
$\vector{G}^1$ and $\vector{G}^2$, and accordingly $\vector{E}_3$ is perpendicular to, both, $\vector{G}^1$ and $\vector{G}^2$.
The components of $\tso{E}$, in the basis $\vector{E}_1$, $\vector{E}_2$, and $\vector{E}_3$,  are denoted with a bar, i.e $\bar{\{ \}}$,
and can be obtained according to following extraction rule
\begin{equation}
\Bar{E}_{ij} = E_{mn} (\vector{E}_i \singlecontraction  \vector{G}^m) (\vector{E}_j \singlecontraction \vector{G}^n).
\label{local_ortho_strain_components}
\end{equation}
The above equation can be simplified for planar components as           
\begin{equation}
\Bar{E}_{\gamma \delta} = E_{\alpha \beta} (\vector{E}_\gamma \singlecontraction  \vector{G}^\alpha) (\vector{E}_\delta \singlecontraction \vector{G}^\beta).
\label{local_ortho_strain}
\end{equation}
By substituting Equation \ref{eab} into \ref{local_ortho_strain}, we have
\begin{align}
\Bar{E}_{\gamma \delta} &= (\varepsilon_{\alpha\beta} + \zeta K_{\alpha\beta}) (\vector{E}_\gamma \singlecontraction  \vector{G}^\alpha) (\vector{E}_\delta \singlecontraction \vector{G}^\beta), \nonumber \\
&= \Bar{\varepsilon}_{\gamma \delta} + \zeta \Bar{K}_{\gamma \delta}
\label{local_ortho_strain_2}
\end{align}
where
\begin{equation}
\Bar{\varepsilon}_{\gamma \delta} = \varepsilon_{\alpha\beta} (\vector{E}_\gamma \singlecontraction  \vector{G}^\alpha) (\vector{E}_\delta \singlecontraction \vector{G}^\beta),
\label{local_ortho_strain_h1}
\end{equation}
and
\begin{equation}
\Bar{K}_{\gamma \delta} = K_{\alpha\beta} (\vector{E}_\gamma \singlecontraction  \vector{G}^\alpha) (\vector{E}_\delta \singlecontraction \vector{G}^\beta).
\label{local_ortho_strain_h2}
\end{equation}
Finally, the relationship between the components of Second Piola-Kirchoff stress and Green-Lagrange strain, in the local orthonormal basis, using Voigt notation, under plane stress conditions is given as
\begin{equation}
\left[\begin{array}{l}
\bar{S}_{11} \\
\bar{S}_{22} \\
\bar{S}_{12}
\end{array}\right]=\dfrac{E}{1-\nu^{2}}\left[\begin{array}{ccc}
1 & \nu & 0 \\
\nu & 1 & 0 \\
0 & 0 & \dfrac{1-\nu}{2}
\end{array}\right]\left[\begin{array}{l}
\bar{E}_{11} \\
\bar{E}_{22} \\
2 \bar{E}_{12}
\end{array}\right].
\end{equation}
\subsection{Principle of virtual work for nonlinear Koiter shell}
For a three dimensional elastic body, the potential energy ($PE$), in terms of strain energy ($SE$), 
and load potential ($LP$), is given as
\begin{equation}
PE = SE - LP
\label{pe-def}
\end{equation}
where, 
\begin{equation}
SE = 
\int_{V_0} W(\tso{E}) \, dV_0,
\label{se-def}
\end{equation}
and
\begin{equation}
LP = \int_{V_0} \rho \, \vector{B} \singlecontraction \vector{u} \, dV_0 + \int_{\Gamma_0} \vector{T} \singlecontraction  \vector{u} \, d\Gamma_0.
\label{lp-def}
\end{equation}
In the above equations, $V_0$ is the reference volume of the body, $\Gamma_0$ denotes the external surface in the reference 
configuration, $\vector{B}$ is the body force per unit volume in the reference configuration, $\vector{T}$ is the 
Piola traction defined as $\vector{T} = \tso{S} \singlecontraction \vector{N}$, where $\vector{N}$ is the unit 
normal on $\Gamma_0$, and $\rho$ is the material density. 
In order to obtain virtual work equation, we set the first variation of the potential energy equal to zero.
$\delta$ is used to denote the variation or a virtual quantity,
using the expressions for $SE$ and $PE$ from 
Equations ~\ref{se-def} and ~\ref{lp-def}, respectively, in Equation ~\ref{pe-def}, we have
\begin{gather}
  \delta PE   =  0  \label{stationarity-pe} \\ 
\implies   \delta SE - \delta LP   = 0 \label{stationarity-pe-1} \\
\implies  \int_{V_0} \tso{S} \doublecontraction \delta \tso{E} \, dV_0 - \int_{V_0} \rho \, \vector{B} \singlecontraction \delta \vector{u} \, dV_0 - 
\int_{\Gamma_0} \vector{T} \singlecontraction \delta \vector{u} \, d\Gamma_0   = 0  \label{stationarity-pe-2} \\
\implies \int_{V_0} \tso{S} \doublecontraction \delta \tso{E} \, dV_0 = \int_{V_0} \rho \, \vector{B} \singlecontraction \delta \vector{u} \, dV_0 + 
\int_{\Gamma_0} \vector{T} \singlecontraction \delta \vector{u} \, d\Gamma_0.  \label{ivw=evw}
\end{gather}
In the above Equations, $\delta \tso{E}$ is the virtual strain, and $\delta \vector{u}$ is virtual displacement.
The left and right hand side terms in Equation \ref{ivw=evw} represent internal and external virtual work, respectively.
Next, we express internal and external virtual work contributions specialized for the Koiter's shell (which are later used for 
extracting nodal point forces in finite element formulation). For the sake of simplicity we will ignore the external body forces from further consideration. 

\subsection{Internal Virtual Work} Using Equations \ref{local_ortho_strain_2}, \ref{local_ortho_strain_h1}, and \ref{local_ortho_strain_h2},
and the plane stress approximation, in the first term on the left hand side of Equation ~\ref{ivw=evw}, 
we have 
\begin{align}
\int_{V_0} \tso{S} \doublecontraction \delta \tso{E} \, dV_0 &= \int_{V_0} \Bar{S}_{\gamma \delta} \,\, \delta \Bar{E}_{\gamma \delta} \, dV_0 \label{ivw_two_term1} \\
&= \int_{V_0} \Bar{S}_{\gamma \delta} \, \left( \delta\Bar{\varepsilon}_{\gamma \delta} + \zeta \delta \Bar{K}_{\gamma \delta} \right) \, dV_0,
\label{ivw_two_term2}
\end{align}
where
\begin{equation}
\delta \Bar{\varepsilon}_{\gamma \delta} = \delta \varepsilon_{\alpha\beta} (\vector{E}_\gamma \singlecontraction  \vector{G}^\alpha) (\vector{E}_\delta \singlecontraction \vector{G}^\beta),
\label{local_ortho_strain_h1_var}
\end{equation}
\begin{equation}
\delta \varepsilon_{\alpha\beta} = \frac{1}{2} \left( \vector{a}_\alpha \singlecontraction \delta \vector{a}_\beta +  \delta \vector{a}_\alpha \singlecontraction \vector{a}_\beta \right),
\label{var_eab}
\end{equation}
\begin{equation}
\delta \Bar{K}_{\gamma \delta} = \delta{K}_{\alpha\beta} (\vector{E}_\gamma \singlecontraction  \vector{G}^\alpha) (\vector{E}_\delta \singlecontraction \vector{G}^\beta),
\label{local_ortho_strain_h2_var}
\end{equation}
and
\begin{align}
\Bar{S}_{\gamma \delta} &= C^{\gamma \delta \alpha \beta} \Bar{E}_{\alpha \beta} \nonumber \\
&= C^{\gamma \delta \alpha \beta} \left( \Bar{\varepsilon}_{\alpha \beta} + \zeta \Bar{K}_{\alpha \beta} \right).
\label{s_bar_gamma_Delta}
\end{align}
Accordingly, the first term on the right hand side in Equation \ref{ivw_two_term2} can simplified using Equation \ref{s_bar_gamma_Delta} as
\begin{align}
\int_{V_0} \Bar{S}_{\gamma \delta} \, \delta\Bar{\varepsilon}_{\gamma \delta} \, dV_0 &= \int_{A_0} \int_{-t/2}^{t/2} \left[ C^{\gamma \delta \alpha \beta} \left( \Bar{\varepsilon}_{\alpha \beta} + \zeta \Bar{K}_{\alpha \beta} \right) \right] \, \delta\Bar{\varepsilon}_{\gamma \delta} \, d\zeta \, dA_0 \nonumber\\
&= \int_{A_0} t \, C^{\gamma \delta \alpha \beta} \Bar{\varepsilon}_{\alpha \beta} \, \delta\Bar{\varepsilon}_{\gamma \delta} \, dA_0 \nonumber \\
&= \int_{A_0} \Bar{n}_{\gamma \delta} \, \delta\Bar{\varepsilon}_{\gamma \delta} \, dA_0,
\end{align}
where
\begin{equation}
\Bar{n}_{\gamma \delta} = t \, C^{\gamma \delta \alpha \beta} \Bar{\varepsilon}_{\alpha \beta},
\label{n_bar_define}
\end{equation}
and through thickness integration in the shell has been carried out. 
The components of $\Bar{n}_{\gamma \delta}$ in Voigt notation are given as
\begin{equation}
\left[\begin{array}{l}
\bar{n}_{11} \\
\bar{n}_{22} \\
\bar{n}_{12}
\end{array}\right]=\dfrac{E t}{1-\nu^{2}}\left[\begin{array}{ccc}
1 & \nu & 0 \\
\nu & 1 & 0 \\
0 & 0 & \dfrac{1-\nu}{2}
\end{array}\right]\left[\begin{array}{l}
\bar{\varepsilon}_{11} \\
\bar{\varepsilon}_{22} \\
2 \bar{\varepsilon}_{12}
\end{array}\right].
\label{eqn:n-bar-calc}
\end{equation}
Similarly, the second term on the right hand side in Equation \ref{ivw_two_term2} simplifies to
\begin{align}
\int_{V_0} \Bar{S}_{\gamma \delta} \zeta \delta \Bar{K}_{\gamma \delta} \, dV_0 &= \int_{A_0} \int_{-t/2}^{t/2} \left[ C^{\gamma \delta \alpha \beta} \left( \Bar{\varepsilon}_{\alpha \beta} + \zeta \Bar{K}_{\alpha \beta} \right) \right] \, \zeta \delta \Bar{K}_{\gamma \delta} \, d\zeta \, dA_0 \nonumber\\
&= \int_{A_0} \int_{-t/2}^{t/2}  C^{\gamma \delta \alpha \beta} \zeta^2 \Bar{K}_{\alpha \beta} \, \delta \Bar{K}_{\gamma \delta} \, d\zeta \, dA_0 \nonumber\\
&= \int_{A_0} \frac{t^3}{12}  C^{\gamma \delta \alpha \beta}  \Bar{K}_{\alpha \beta} \, \delta \Bar{K}_{\gamma \delta} \, dA_0 \nonumber\\
&= \int_{A_0} \Bar{m}_{\gamma \delta} \, \delta \Bar{K}_{\gamma \delta} \, dA_0,
\end{align}
where using Equations \ref{Kab} and \ref{bab_def}, along with some algebraic simplifications, we have
\begin{align}
\delta K_{\alpha\beta} &= - \delta b_{\alpha\beta} \nonumber \\
&= - \Big[ \delta \vector{a}_{\alpha, \beta} - \left( \vector{a}^{\gamma} \singlecontraction \vector{a}_{\alpha, \beta} \right) \delta \vector{a}_\gamma \Big] \singlecontraction \athree,
\label{varkappaab}
\end{align}
\begin{equation}
\delta \Bar{K}_{\gamma \delta} = - (\vector{E}_\gamma \singlecontraction  \vector{G}^\alpha) (\vector{E}_\delta \singlecontraction \vector{G}^\beta) \Big[ \delta \vector{a}_{\alpha, \beta} - \left( \vector{a}^{\gamma} \singlecontraction \vector{a}_{\alpha, \beta} \right) \delta \vector{a}_\gamma \Big] \singlecontraction \athree,
\label{kappa_bar_gamma_delta}
\end{equation}
\begin{equation}
\Bar{m}_{\gamma \delta} = \frac{t^3}{12}  C^{\gamma \delta \alpha \beta}  \Bar{K}_{\alpha \beta}.
\label{m_bar_gamma_delta_define}
\end{equation}
The components $\Bar{m}_{\gamma \delta}$ in Voigt notation are given as
\begin{equation}
\left[\begin{array}{l}
\bar{m}_{11} \\
\bar{m}_{22} \\
\bar{m}_{12}
\end{array}\right]=\dfrac{E t^3}{12(1-\nu^{2})}\left[\begin{array}{ccc}
1 & \nu & 0 \\
\nu & 1 & 0 \\
0 & 0 & \dfrac{1-\nu}{2}
\end{array}\right]\left[\begin{array}{l}
\bar{K}_{11} \\
\bar{K}_{22} \\
2 \bar{K}_{12}
\end{array}\right].
\label{eqn:m-bar-calc}
\end{equation}
Thus, the final expression for the internal virtual work is given as 
\begin{equation}
\int_{V_0} \tso{S} \doublecontraction \delta \tso{E} \, dV_0 = \int_{A_0} \left( \Bar{n}_{\gamma \delta} \, \delta \Bar{\epsilon}_{\gamma \delta} + \Bar{m}_{\gamma \delta} \, \delta \Bar{K}_{\gamma \delta} \right) dA_{0},
\label{ivw}
\end{equation}
where $\Bar{n}_{\gamma \delta}$, $\delta \Bar{\epsilon}_{\gamma \delta}$, $\Bar{m}_{\gamma \delta}$, and $\delta \Bar{K}_{\gamma \delta}$ are given in the Equations \ref{n_bar_define}, \ref{local_ortho_strain_h1_var}, \ref{m_bar_gamma_delta_define}, and \ref{kappa_bar_gamma_delta}, respectively. Correspondingly, we identify the membrane energy ($E_M$), and bending energy ($E_B$) for this model as
\begin{align}
E_M &= \frac{1}{2} \int_{A_0} \Bar{n}_{\gamma \delta} \Bar{\varepsilon}_{\gamma \delta} \, dA_0,
\label{membrane_energy_expr}
\end{align}
and 
\begin{align}
E_B &= \frac{1}{2} \int_{A_0} \Bar{m}_{\gamma \delta} \Bar{K}_{\gamma \delta} \, dA_0.
\label{bending_energy_expr}
\end{align}

\subsection{External Virtual  Work}
We divide the external surface of the shell $\Gamma_0$ into lateral surface $\Omega_0$, and top and bottom surfaces ${R}^+_0$ and 
${R}^-_{0}$, respectively, i.e., $\Gamma_0$ = 	$ \Omega_0 \cup {R}^+_0 \cup {R}^-_{0}$. For sake of simplicity, without any loss of generality, we assume that only $\Omega_0$ is loaded. $\Omega_0$ is further decomposed into ${\Omega}_0^N$ and  
${\Omega}_0^U$ where the traction and displacement boundary conditions are specified, respectively. Accordingly, 
we further decompose the lateral surface as a product of corresponding edge boundary ($\omega_0$) and through thickness, i.e., 
${\Omega}_0^N$ = $\left[ -\frac{t}{2},\frac{t}{2}\right] \times {\omega}_0^N $, 
and ${\Omega}_0^U$ = $\left[ -\frac{t}{2},\frac{t}{2}\right] \times {\omega}_0^U$.
The virtual work due to 
Piola traction on ${\Omega}_0^N$ is given as
\begin{align}
\int_{{\Omega}_0^N} \vector{T} \singlecontraction \delta \vector{u} \, d{\Omega}_0^N = \int_{{\Omega}_0^N} \vector{T} \singlecontraction \delta \vector{x} (\theta^1, \theta^2, \zeta) \, d{\Omega}_0^N,
\label{boundary-work-shell}
\end{align}
where the variation in displacement is obtained by taking variation in the deformed position vector $\vector{x}(\theta^1, \theta^2, \zeta)$.
Using Equation ~\ref{kinematic_assumption_def} we have 
\begin{equation}
\delta \vector{x}(\theta^1, \theta^2, \zeta) = \delta \vector{x}(\theta^1, \theta^2) + \zeta \delta \athree,
\label{variation_def_location}
\end{equation}
substituting Equation ~\ref{variation_def_location} in Equation ~\ref{boundary-work-shell}, we derive
\begin{align}
\int_{{\Omega}_0^N} \vector{T} \singlecontraction \delta \vector{x} (\theta^1, \theta^2, \zeta) \, d{\Omega}_0^N = & 
\int_{{\Omega}_0^N} \vector{T} \singlecontraction (\delta \vector{x}(\theta^1, \theta^2) + \zeta \delta \athree) \, d{\Omega}_0^N \\
= & \int_{{\Omega}_0^N} \vector{T} \singlecontraction \delta \vector{x}(\theta^1, \theta^2) \, d{\Omega}_0^N +  \int_{{\Omega}_0^N} \vector{T} \zeta  \singlecontraction \delta \athree \, d{\Omega}_0^N \\
= & \int_{\omega_0^N} \left[ \int_{-t/2}^{t/2} \vector{T} d \zeta \right] \singlecontraction \delta \vector{x}(\theta^1, \theta^2) \, d \omega_0^N + 
\int_{\omega_0^N} \left[ \int_{-t/2}^{t/2} \vector{T} d \zeta \right] \singlecontraction \delta \athree   \, d \omega_0^N.
\label{boundary-work-shell-2}
\end{align}
We define $\vector{N}_o$ = $\int_{-t/2}^{t/2} \vector{T} d \zeta$ and $\vector{M}_o$ = $\int_{-t/2}^{t/2} \vector{T} d \zeta $
as the boundary membrane forces and bending moments, respectively. We note that the imposition of displacement boundary conditions requires  
specification of the displacement field $\vector{u}$, and its normal derivative $\vector{u}_{,{\nu}}$ on  $\omega_0^U$. 

\section{Penalty Formulations}
\label{sec:penalty-formulation}
For the examples chosen in this study, the given boundary data for the edge displacements, denoted by $\bar{\vector{u}}$, 
is directly incorporated into the location of the control points, and the slopes ($\bar{\vector{u}}_{,\nu}$) are imposed using a penalty formulation.  
Similarly, for imposing point displacement 
on the shell midsurface, and enforcing $C^1$ continuity between two mating external edges of a NURBS geometry, where only $C^0$ continuity 
is available, we use the penalty method. Although we do not provide a detailed numerical analysis in this work, we note that the penalty term provides a coercive, lower weakly semi-continuous addition to the energy functional, in the weak topology of appropriate Sobolev space. 
We also suppress the explicit notation for mesh-dependent penalty parameter. 

\subsection{Enforcing slope-based boundary conditions}
\label{known_grad_u_penalty_section}
To enforce slope-based Dirchlet boundary conditions, we add the following penalty term $S_G$ to the total potential energy $PE$ (Equation ~\ref{pe-def}) as
\begin{equation}
S_{G} = \int_{ \omega_0^U} K_{G} \left( \vector{u}_{,\nu} - \vector{$\bar{u}$}_{,\nu} \right) \singlecontraction \left( \vector{u}_{,\nu} - \vector{$\bar{u}$}_{,\nu} \right) d\omega_0^U,
\label{given_disp_grad_penalty}
\end{equation}
where $K_G$ is the penalty parameter. Accordingly, its variation is also included in the stationarity condition of the total potential energy (Equation ~\ref{stationarity-pe}).
The variation of $S_G$ is given as 
\begin{equation}
\delta S_G = \int_{\omega_0^U} 2 K_G \left( \vector{u}_{,\nu} - \vector{$\bar{u}$}_{,\nu} \right) \singlecontraction \delta \vector{u}_{,\nu} \,\, d\omega_0^U.
\label{given_disp_grad_penalty_var}
\end{equation}

\subsection{Point displacements}
Control points in NURBS-based discretization do not necessarily lie on the material surface; thus, in order to impose displacement based boundary conditions on any material point on the midsurface shell, we add a displacement penalty term $U_p$ to the potential energy expression in Equation ~\ref{pe-def}. 
The displacement penalty is given as 
\begin{equation}
U_p = K_u \left( \vector{u} - \bar{\vector{u}} \right) \singlecontraction \left( \vector{u} - \bar{\vector{u}} \right),
\label{disp_penalty}
\end{equation}
where $K_u$ is the penalty parameter, and $\bar{\vector{u}}$ is the specified displacement at a particular point. The variation of 
$U_p$ is also included in the stationarity condition of potential energy, Equation ~\ref{stationarity-pe}, and is given 
as
\begin{equation}
\delta U_p = 2 K_u \left( \vector{u} - \bar{\vector{u}} \right) \singlecontraction  \delta \vector{u}.
\label{disp_penalty_var}
\end{equation}

\subsection{$C^1$ continuity along mating exterior edges} 
\label{disp_grad_penaly_section}
When two exterior edges of a single patch NURBS surface, represented the open vectors, coincide and form a common edge (denoted as $E$), only $C^0$ continuity can be expected at the common edge.
In order to achieve $C^1$ continuity at the common edge, we can enforce equality between the derivatives of the displacements, normal to the common edge, computed on 
both sides of the common edge by adding a penalty term based on their differences in the total potential energy. We construct such a penalty term $S_p$ as 
\begin{equation}
S_p = \int_{E} K_s \left( \vector{u}_{,\Lambda}^{+} - \vector{u}_{,\Lambda}^{-} \right) \singlecontraction \left( \vector{u}_{,\Lambda}^{+} - \vector{u}_{,\Lambda}^{-} \right) 
d E,
\label{disp_grad_penalty}
\end{equation}
where $K_s$ is the penalty parameter, and $\vector{u}_{,\Lambda}^{+}$ and $\vector{u}_{,\Lambda}^{-}$ are displacement derivatives in the direction normal to the edge $E$, the normal direction in curvilinear coordinates is denoted by $\Lambda$.
The variation of $S_p$  required in stationarity of the potential energy is given as 
\begin{equation}
\delta S_p = \int_{E} 2 K_s \left( \vector{u}_{,\Lambda}^{+} - \vector{u}_{,\Lambda}^{-} \right) \singlecontraction \left( \delta \vector{u}_{,\Lambda}^{+} - \delta \vector{u}_{,\Lambda}^{-} \right) dE.
\label{disp_grad_penalty_var}
\end{equation}

\section{NURBS-based Isogeometric Analysis}
\label{sec:nurbs}
We employ Isogeometric Analysis, and discretize the midsurface of the shell in the reference and deformed configurations using NURBS. 
The position vector $\vector{S}_{ref}(\xi, \eta)$ of any point in the reference configuration, parameterized by 
NURBS coordinates ($\xi$, $\eta$), is given by interpolation of the position vectors of the control points 
$\vector{P}_{i,j}^{ref}$ as

\begin{equation}
\vector{S}_{ref}(\xi, \eta) = \sum_{i=1}^{n} \sum_{j=1}^{m} R_{i,j}^{p,q}(\xi, \eta) \,\, \vector{P}_{i,j}^{ref},
\label{ref_nurbs_equation}
\end{equation}
where $R_{i,j}^{p,q}(\xi, \eta)$ represent the NURBS shape functions defined in terms of rational basis functions
as 
\begin{equation}
R_{i,j}^{p,q} (\xi, \eta) = \frac{N_{i,p} (\xi) M_{j,q} (\eta) W_{i,j}}{ \sum_{a=1}^{n} \sum_{b=1}^{m} N_{a,p} (\xi) M_{b,q} (\eta) W_{a,b} },
\label{nurbs-function}
\end{equation}
with $W_{i,j}$ as the weight of the control point indexed by $i$ and $j$ on an $n \times m$ two dimensional NURBS parametric grid. The $W_{i,j}s$ are taken to be $1.0$ here for flat geometries. We denote by $N$ the total number of control points, which is equal to $n\times m$.
$N_{i, p}(\xi)$ and $M_{j, q}(\eta)$ 
are the B-spline basis functions of degrees $p$ and $q$, respectively, and are associated with the open knot vectors 
$\Xi_1= \{\xi_{1},.....,  \xi_{n+p+1} \}^{T}$ and $\Xi_2 = \{\eta_{1},.....,\eta_{m+p+1}\}^{T}$, respectively. 
The $\Xi_1$ and $\Xi_2$ are known as the ``open knot vectors''. The end knots are repeated $p+1$ and $q+1$ times. $N_{i, p}(\xi)$ and $M_{j, q}(\eta)$ are defined based on the Cox--de Boor recursion formula as follows

\begin{equation}
N_{i,p}(\xi) = 
\begin{cases}
 p=0 &
 
\begin{cases}
    1& \text{if} \indent \xi_i \leq \xi < \xi_{i+1}, \\
    0              & \text{otherwise.}
\end{cases} \\\\

 p>0 & 
\frac{\xi - \xi_i}{\xi_{i+p} - \xi_i} N_{i, p-1}(\xi) + \frac{\xi_{i+p+1} - \xi}{\xi_{i+p+1} - \xi_{i+1}} N_{i+1, p-1}(\xi) 

\end{cases}
\end{equation}
and,
\begin{equation}
M_{j,q}(\eta) = 
\begin{cases}
 q=0 &
 
\begin{cases}
    1& \text{if} \indent \eta_j \leq \eta < \eta_{j+1}, \\
    0              & \text{otherwise.}
\end{cases} \\\\

 q>0 & 
\frac{\eta - \eta_j}{\eta_{j+q} - \eta_j} M_{j, q-1}(\eta) + \frac{\eta_{j+q+1} - \eta}{\eta_{j+q+1} - \eta_{j+1}} M_{j+1, q-1}(\eta).

\end{cases}
\end{equation}
Similarly, any physical point on the deformed surface is defined as 
\begin{equation}
\vector{S}_{def}(\xi, \eta) = \sum_{i=1}^{n} \sum_{j=1}^{m} R_{i,j}^{p,q}(\xi, \eta) \,\, \vector{P}_{i,j}^{def},
\label{def_nurbs_equation}
\end{equation}
where \vector{P}$_{i,j}^{def}$ are the position vectors of the control points in the deformed configuration. 
For the simulations carried out in this work, we have used p=q=3.

\subsection{Finite Element Discretization}
\label{sec:FEA-discretization}
In this section, we discuss the finite element discretization using NURBS. We use a Bubnov--Galerkin formulation, by using 
the same NURBS basis functions for shape functions and the trial functions. For the sake of simplicity we present a straight-forward 
discretization scheme by using global matrices; however, the implementation is carried out efficiently by taking into account 
the sparsity of different matrices, and economizing the storage. The displacements of the control points are arranged in a column vector $\left [ \vector{u} \right ]_{3N \times 1}$ of size $3N \times 1$, since there are a total of $N$ control points, and each control point has three degrees of freedom.
The NURBS shape functions, from Equation \ref{nurbs-function}, which interpolate the position vectors of the control points, are arranged in  the matrix $\left [ R \right ]_{3\times 3N}$ of size $3 \times 3N$, 
\begin{equation}
\left[\begin{array}{l}
R
\end{array}\right]_{3 \times 3N} = \left[\begin{array}{cccccccccccc}
R_1 & 0   & 0   & R_2 & 0   & 0   & . & . & . & R_N & 0   & 0   \\
0   & R_1 & 0   &   0 & R_2 & 0   & . & . & . &   0 & R_N & 0   \\
0   & 0   & R_1 &   0 &   0 & R_2 & . & . & . &   0 & 0   & R_N \\
\end{array}\right]_{_{3 \times 3N}},
\label{matrix_R_components}
\end{equation}
only $(p+1)\times(q+1)$ basis functions (and control points) find support within a particular element. The 
first order and second order derivatives of the shape functions with respect to curvilinear coordinates are given as 
\begin{equation}
\arraycolsep=3.5pt
\left[\begin{array}{l}
\dfrac{\partial R}{\partial \theta^{\alpha}}
\end{array}\right]_{3 \times 3N} = \left[\begin{array}{cccccccccccc}
\dfrac{\partial R_1}{\partial \theta^{\alpha}} & 0   & 0   & \dfrac{\partial R_2}{\partial \theta^{\alpha}} & 0   & 0   & . & . & . & \dfrac{\partial R_N}{\partial \theta^{\alpha}}& 0   & 0   \\
0   & \dfrac{\partial R_1}{\partial \theta^{\alpha}} & 0   &   0 & \dfrac{\partial R_2}{\partial \theta^{\alpha}} & 0   & . & . & . &   0 & \dfrac{\partial R_N}{\partial \theta^{\alpha}} & 0   \\
0   & 0   & \dfrac{\partial R_1}{\partial \theta^{\alpha}} &   0 &   0 & \dfrac{\partial R_2}{\partial \theta^{\alpha}} & . & . & . &   0 & 0   & \dfrac{\partial R_N}{\partial \theta^{\alpha}} \\
\end{array}\right]_{_{3 \times 3N}},
\label{matrix_dR_dtheta_alpha_components}
\end{equation}
and
\begin{equation}
\arraycolsep=0.8pt
\left[\begin{array}{l}
\dfrac{\partial^2 R}{\partial \theta^{\alpha} \partial \theta^{\beta}}
\end{array}\right]_{3 \times 3N} = \left[\begin{array}{cccccccccccc}
\frac{\partial^2 R_1}{\partial \theta^{\alpha} \partial \theta^{\beta}} & 0   & 0   & \frac{\partial^2 R_2}{\partial \theta^{\alpha} \partial \theta^{\beta}} & 0   & 0   & . & . & . & \frac{\partial^2 R_N}{\partial \theta^{\alpha} \partial \theta^{\beta}} & 0   & 0   \\
0   & \frac{\partial^2 R_1}{\partial \theta^{\alpha} \partial \theta^{\beta}} & 0   &   0 & \frac{\partial^2 R_2}{\partial \theta^{\alpha} \partial \theta^{\beta}} & 0   & . & . & . &   0 & \frac{\partial^2 R_N}{\partial \theta^{\alpha} \partial \theta^{\beta}} & 0   \\
0   & 0   & \frac{\partial^2 R_1}{\partial \theta^{\alpha} \partial \theta^{\beta}}&   0 &   0 & \frac{\partial^2 R_2}{\partial \theta^{\alpha} \partial \theta^{\beta}} & . & . & . &   0 & 0   & \frac{\partial^2 R_N}{\partial \theta^{\alpha} \partial \theta^{\beta}} \\
\end{array}\right]_{_{3 \times 3N}}.
\label{matrix_d2R_dtheta_alpha_beta_components}
\end{equation}

The displacement vector $\vector{u}$ at any parametric point $(\xi, \eta)$ is obtained using the shape function matrix at $(\xi, \eta)$, and displacement vector of control points as
\begin{align}
 \vector{u} & = \left [ R \right ]_{3\times 3N} \left [ \vector{u} \right ]_{3N \times 1},
\label{u_interp}
\end{align}
similarly the variation of the displacement field is given as 
\begin{align}
  \delta \vector{u} & = \left [ R \right ]_{3\times 3N} \left [ \delta  \vector{u} \right ]_{3N \times 1}.
\label{var_u_interp}
\end{align}
Next, we discuss the discretization of various terms in the virtual work equation, including discretization of the penalty terms:

\begin{enumerate}

\item The internal force vector due to internal stresses is obatined by introducing isogeometric discretization in the internal virtual work (Equation \ref{ivw}) as 
\begin{equation}
\left[ \vector{F}_{\text{s}} \right]^{\top}_{_{3N \times 1}} \left[ \delta \vector{u} \right]_{_{3N \times 1}}=
\int_{A_0} \left( \Bar{n}_{\gamma \delta} \, \delta \Bar{\epsilon}_{\gamma \delta} + \Bar{m}_{\gamma \delta} \, \delta \Bar{K}_{\gamma \delta} \right) dA_{0} = \sum_{i=1}^{N_e} \int_{A_0^i} \left( \Bar{n}_{\gamma \delta} \, \delta \Bar{\epsilon}_{\gamma \delta} + \Bar{m}_{\gamma \delta} \, \delta \Bar{K}_{\gamma \delta} \right)^i dA_{0}^i,
\label{internal_virtual_work_disc}
\end{equation}
where $\left[ \vector{F}_{\text{s}} \right]^{\top}_{_{3N \times 1}}$ is the nodal point force vector, 
$\left[ \delta \vector{u} \right]_{_{3N \times 1}}$ is a virtual displacement vector of size $3N \times 1$, the integral over the reference midsurface is broken into sum of integrals over the NURBS elements (spanning the shell midsurface in the reference configuration), with the index $i$ indicating the element number. 
The computation of $\Bar{n}_{\gamma \delta}$ and $\Bar{m}_{\gamma \delta}$ is specified in Equations ~\ref{eqn:n-bar-calc}
and ~\ref{eqn:m-bar-calc}, respectively, and the variations in membrane strains, and curvatures, following Equations ~\ref{local_ortho_strain_h1_var}, ~\ref{var_eab}, ~\ref{local_ortho_strain_h2_var}, and ~\ref{kappa_bar_gamma_delta}, can be expressed as

\begin{align}
\delta \Bar{\varepsilon}_{\gamma \delta} &= \frac{1}{2} \left[\vector{E}_\gamma \right]^\top_{_{3 \times 1}} \left[\vector{G}^\alpha \right]_{_{3 \times 1}} \left[ \vector{E}_\delta \right]^\top_{_{3 \times 1}} \left[\vector{G}^\beta \right]_{_{3 \times 1}} \left( \left[ \vector{a}_\alpha \right]^\top_{_{3 \times 1}} \left[ \delta \vector{a}_\beta \right]_{_{3 \times 1}} +  \left[ \vector{a}_\beta \right]^\top_{_{3 \times 1}} \left[\delta \vector{a}_\alpha \right]_{_{3 \times 1}} \right)\nonumber \\
&= \frac{1}{2} \left[ \vector{E}_\gamma \right]^\top_{_{3 \times 1}} \left[ \vector{G}^\alpha \right]_{_{3 \times 1}} \left[ \vector{E}_\delta \right]^\top_{_{3 \times 1}} \left[ \vector{G}^\beta \right]_{_{3 \times 1}} \left( \left[ \vector{a}_\alpha \right]^\top_{_{3 \times 1}} \left[ \frac{\partial R}{\partial \theta^{\beta}} \right]_{_{3 \times 3N}} + \left[ \vector{a}_\beta \right]^\top_{_{3 \times 1}} \left[ \frac{\partial R}{\partial \theta^{\alpha}} \right]_{_{3 \times 3N}} \right) \left[ \delta \vector{u} \right]_{_{3N \times 1}}, 
\label{disc_delta_ebar}
\end{align}
and
\begin{align}
\delta \Bar{K}_{\gamma \delta} &= \left[ \vector{E}_\gamma \right]^\top_{_{3 \times 1}} \left[\vector{G}^\alpha \right]_{_{3 \times 1}} \left[ \vector{E}_\delta \right]^\top_{_{3 \times 1}} \left[\vector{G}^\beta \right]_{_{3 \times 1}} \left( \left[ \vector{a}^{\gamma} \right]^\top_{_{3 \times 1}}  \left[\vector{a}_{\alpha, \beta} \right]_{_{3 \times 1}} \left[ \vector{a}_3 \right]^\top_{_{3 \times 1}} \left[ \delta \vector{a}_{\gamma} \right]_{_{3 \times 1}} - \left[ \vector{a}_3 \right]^\top_{_{3 \times 1}} \left[ \delta \vector{a}_{\alpha, \beta} \right]_{_{3 \times 1}} \right) \nonumber  \\
&= \left[ \vector{E}_\gamma \right]^\top_{_{3 \times 1}} \left[ \vector{G}^\alpha \right]_{_{3 \times 1}} \left[ \vector{E}_\delta \right]^\top_{_{3 \times 1}} \left[ \vector{G}^\beta \right]_{_{3 \times 1}} \Bigg( \left[ \vector{a}^{\gamma} \right]^\top_{_{3 \times 1}}  \left[ \vector{a}_{\alpha, \beta} \right]_{_{3 \times 1}} \left[ \vector{a}_3 \right]^\top_{_{3 \times 1}} \left[ \frac{\partial R}{\partial \theta^{\gamma}} \right]_{_{3 \times 3N}} \nonumber  \\ 
& \quad\quad\quad\quad\quad\quad\quad\quad\quad\quad\quad\quad\quad\quad\quad\quad - \left[ \vector{a}_3 \right]^\top_{_{3 \times 1}} \left[ \frac{\partial^2 R}{\partial \theta^{\alpha} \partial \theta^{\beta}} \right]_{_{3 \times 3N}}  \Bigg) \left[ \delta \vector{u} \right]_{_{3N \times 1}},
\label{disc_delta_kbar}
\end{align}
By substituting $\delta \Bar{\varepsilon}_{\gamma \delta}$ and $\delta \Bar{K}_{\gamma \delta}$ in the right hand side of Equation \ref{internal_virtual_work_disc}, we can extract the ``nodal'' force vector $ \left[ \vector{F}_{\text{s}} \right]_{_{3N \times 1}}$ (due to internal stresses) as follows
\begin{align}
\left[ \vector{F}_{\text{s}} \right]^{\top}_{_{3N \times 1}} \left[ \delta \vector{u} \right]_{_{3N \times 1}}=
\sum_{i=1}^{N_e} \int_{A_0^i} \left( \Bar{n}_{\gamma \delta} \, \delta \Bar{\epsilon}_{\gamma \delta} + \Bar{m}_{\gamma \delta} \, \delta \Bar{K}_{\gamma \delta} \right)^i dA_{0}^i = \sum_{i=1}^{N_e} \int_{A_0^i} \Bigg[ \frac{1}{2} \Bar{n}_{\gamma \delta} \, \left[ \vector{E}_\gamma \right]^\top_{_{3 \times 1}} \left[ \vector{G}^\alpha \right]_{_{3 \times 1}} \left[ \vector{E}_\delta \right]^\top_{_{3 \times 1}} \left[ \vector{G}^\beta \right]_{_{3 \times 1}} \nonumber \\
\left( \left[ \vector{a}_\alpha \right]^\top_{_{3 \times 1}} \left[ \frac{\partial R}{\partial \theta^{\beta}} \right]_{_{3 \times 3N}} + \left[ \vector{a}_\beta \right]^\top_{_{3 \times 1}} \left[ \frac{\partial R}{\partial \theta^{\alpha}} \right]_{_{3 \times 3N}} \right) + \Bar{m}_{\gamma \delta} \, \left[ \vector{E}_\gamma \right]^\top_{_{3 \times 1}} \left[ \vector{G}^\alpha \right]_{_{3 \times 1}} \left[ \vector{E}_\delta \right]^\top_{_{3 \times 1}} \left[ \vector{G}^\beta \right]_{_{3 \times 1}} \nonumber \\
\Bigg( \left[ \vector{a}^{\gamma} \right]^\top_{_{3 \times 1}} \left[ \vector{a}_{\alpha, \beta} \right]_{_{3 \times 1}} \left[ \vector{a}_3 \right]^\top_{_{3 \times 1}} \left[ \frac{\partial R}{\partial \theta^{\gamma}} \right]_{_{3 \times 3N}} - \left[ \vector{a}_3 \right]^\top_{_{3 \times 1}} \left[ \frac{\partial^2 R}{\partial \theta^{\alpha} \partial \theta^{\beta}} \right]_{_{3 \times 3N}}  \Bigg) \Bigg]^i  dA_{0}^i \left[ \delta \vector{u} \right]_{_{3N \times 1}}.
\label{final_eqn_f_int}
\end{align}

\item To obtain the force vector $\left[ \vector{F}_{\text{G}} \right]^{\top}_{_{3N \times 1}}$
corresponding to slope-based boundary conditions on the control points, we introduce discretization in Equation ~\ref{given_disp_grad_penalty_var}, 
such that 
\begin{equation}
\left[ \vector{F}_{\text{G}} \right]^{\top}_{_{3N \times 1}}  \left[ \delta \vector{u} \right]_{_{3N \times 1}}
= 2 K_G  \Big(\left [ \vector{u}\right]^{\top}_{_{3N \times 1}} \int_{\omega_0^U} [ R_{,\nu} ]^\top_{_{3 \times 3N}} \, [R_{,\nu}]_{_{3 \times 3N}} \, d\omega_0^U  -  [\vector{$\bar{u}$}_{,\nu}]^{\top}_{_{3N \times 1}} \int_{\omega_0^U} R_{,\nu} \, d \omega_0^U \Big)  {[\delta \vector{u}]_{_{3N \times 1}}}.
\label{F_g}
\end{equation}

\item The force vector corresponding to imposition of point displacement is obtained by introducing discretization in Equation ~\ref{disp_penalty_var} as
\begin{equation}
\left[ \vector{F}_{\text{p}} \right]^{\top}_{_{3N \times 1}}  \left[ \delta \vector{u} \right]_{_{3N \times 1}}
=  2 K_u \left( [R]_{3 \times 3N} \, [\vector{u}]_{3N \times 1} - [\bar{\vector{u}}]_{3 \times 1} \right)^{\top} [R]_{3 \times 3N} \, [\delta \vector{u}]_{3N \times 1}.
\label{force-point-disp}
\end{equation}

\item Finally, the force vector arising from enforcing the $C^1$ continuity along the mating edges is obtained by introducing discretization in Equation 
~\ref{disp_grad_penalty_var} as follows

\begin{align}
[\vector{F}_{\text{sp}}]^\top_{_{3N \times 1}} [\delta \vector{u}]_{_{3N \times 1}} &= 2 K_s{[\vector{u}]}^\top_{_{3N \times 1}}  \int_{ E}   \left([R_{,\eta}^{+}]_{_{3 \times 3N}} - [R_{,\eta}^{-}]_{_{3 \times 3N}} \right)^\top \left( [R_{,\eta}]^{+}_{_{3 \times 3N}} - [R_{,\eta}^{-}]_{_{3 \times 3N}} \right) d E  \;  [\delta\vector{u}]_{_{3N \times 1}}. \nonumber \\
& 
\label{disp_grad_penalty_var2}
\end{align}
The line integral in equation \ref{disp_grad_penalty_var2} is performed with the assumption that the number of NURBS elements at the common edge are equal.

\end{enumerate}

Next we discuss the computations of various quantities $\left[ \vector{E}_\gamma \right]_{_{3 \times 1}}$, $\left[ \vector{G}^\alpha \right]_{_{3 \times 1}}$, $\left[ \vector{a}_\alpha \right]_{_{3 \times 1}}$, $\left[ \vector{a}^{\gamma} \right]_{3 \times 1}$, $\left[ \frac{\partial R}{\partial \theta^{\beta}} \right]_{_{3 \times 3N}}$, $\left[ \vector{a}_{\alpha, \beta} \right]_{_{3 \times 1}}$, $\left[ \vector{a}_3 \right]_{_{3 \times 1}}$,
$\left[ \frac{\partial R}{\partial \nu} \right]_{_{3 \times 3N}}$, and 
$\left[ \frac{\partial^2 R}{\partial \theta^{\alpha} \partial \theta^{\beta}} \right]_{_{3 \times 3N}}$
given in Equations ~\ref{final_eqn_f_int},  ~\ref{F_g}, 
~\ref{force-point-disp}, and ~\ref{disp_grad_penalty_var2} to estimate the respective force vectors. 
Numerical quadrature scheme is used to evaluate these quantities at any Gauss point. 

\subsection{Numerical computations}
We use spatial finite differences with second order accuracy to compute various quantities numerically within the element domain, and at the edges. 
$\left[ \vector{E}_\gamma \right]_{_{3 \times 1}}$, $\left[ \vector{G}^\alpha \right]_{_{3 \times 1}}$, $\left[ \vector{a}_\alpha \right]_{_{3 \times 1}}$, $\left[ \vector{a}^{\gamma} \right]_{3 \times 1}$, $\left[ \frac{\partial R}{\partial \theta^{\beta}} \right]_{_{3 \times 3N}}$, $\left[ \vector{a}_{\alpha, \beta} \right]_{_{3 \times 1}}$, $\left[ \vector{a}_3 \right]_{_{3 \times 1}}$, and $\left[ \frac{\partial^2 R}{\partial \theta^{\alpha} \partial \theta^{\beta}} \right]_{_{3 \times 3N}}$ are computed within the element domain,
and $\left[ \frac{\partial R}{\partial \nu} \right]_{_{3 \times 3N}}$ is computed on the boundary edge. We first discuss the computations within the element domain, and then at the edge. 

\begin{figure}[ht!]
    \centering
    \includegraphics[scale=1.5]{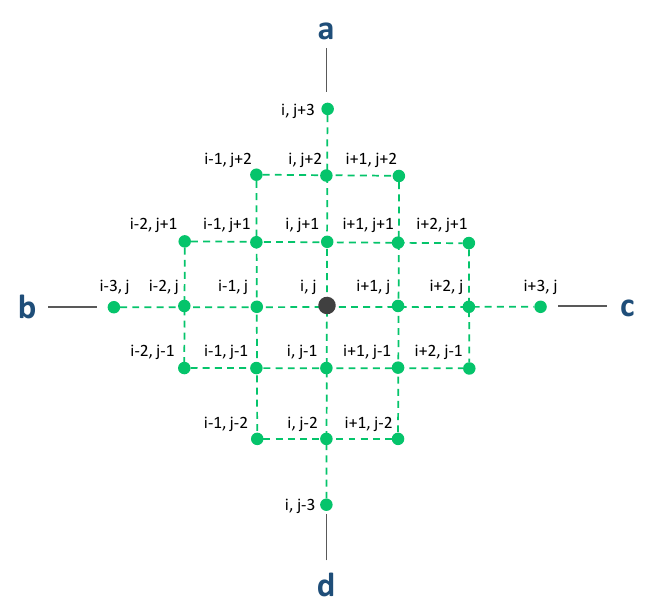}
    \caption{A schematic of the spatial network of 25 points, around the Gauss point $(i,j)$, in the NURBS space.}
    \label{25-points-network}
\end{figure}

\subsubsection{Computations within the element domain} 
\label{inside_element_computations}

A discrete network of 25 points is initialized at any Gauss point, indexed by $(i, j)$,  in the NURBS space, at incremental distances of
$\varepsilon = 10^{-5}$ in $\xi$ and $\eta$ parametric directions. The complete network is shown in Figure \ref{25-points-network}, where 
the network points, in $\xi$ and $\eta$ parametric directions, are correspondingly indexed by $(i+g, j+h)$, with $g$, $h$ $\in$ $\mathbb{Z}$. 
All network points are mapped to, both, 
reference and deformed configurations (according to Equations \ref{ref_nurbs_equation} and \ref{def_nurbs_equation}). 
Central difference is used throughout the approximations within the element domain.

\begin{enumerate}

\item To obtain reference covariant vectors $\left[(\vector{G}_1)_{i,j}\right]_{_{3 \times 1}}$ and $\left[(\vector{G}_2)_{i,j}\right]_{_{3 \times 1}}$, at the index $(i,j)$, 
we use the reference position vectors of the network points  $\left[\vector{X}\right]_{_{3 \times 1}}$ at neighbouring indexes as follows
\begin{equation}
\left[(\vector{G}_1)_{i,j}\right]_{_{3 \times 1}} = \frac{\left[\vector{X}_{i+1, j}\right]_{_{3 \times 1}} - \left[\vector{X}_{i-1, j}\right]_{_{3 \times 1}} }{ ||\left[\vector{X}_{i+1, j}\right]_{_{3 \times 1}} - \left[\vector{X}_{i-1, j}\right]_{_{3 \times 1}} || },    
\end{equation}
and
\begin{equation}
\left[(\vector{G}_2)_{i,j}\right]_{_{3 \times 1}} = \frac{\left[\vector{X}_{i, j+1}\right]_{_{3 \times 1}} - \left[\vector{X}_{i, j-1}\right]_{_{3 \times 1}} }{ ||\left[\vector{X}_{i, j+1}\right]_{_{3 \times 1}} - \left[\vector{X}_{i, j-1}\right]_{_{3 \times 1}} || }.
\end{equation}

\item As discussed in Section ~\ref{shell-kinematics-derivation}, the contravariant vectors $\left[(\vector{G}^1)_{i,j} \right]_{_{3 \times 1}}$ and 
$\left[(\vector{G}^2)_{i,j}\right]_{_{3 \times 1}}$ lie in the plane formed by $\left[(\vector{G}_1)_{i,j}\right]_{_{3 \times 1}}$, and $\left[(\vector{G}_2)_{i,j}\right]_{_{3 \times 1}}$, and can be 
expressed as a linear combination of $\left[(\vector{G}_1)_{i,j}\right]_{_{3 \times 1}}$ and $\left[(\vector{G}_2)_{i,j}\right]_{_{3 \times 1}}$. Accordingly,

\begin{equation}
\left[\vector{G}^1\right]_{_{3 \times 1}} = \frac{1}{b^2 - ac} \left( b \, \left[\vector{G}_2\right]_{_{3 \times 1}} - c \, \left[\vector{G}_1\right]_{_{3 \times 1}} \right),
\end{equation}
and
\begin{equation}
\left[\vector{G}^2\right]_{_{3 \times 1}} = \frac{1}{b^2 - ac} \left( b \, \left[\vector{G}_1\right]_{_{3 \times 1}} - a \, \left[\vector{G}_2\right]_{_{3 \times 1}} \right),
\end{equation}
where the scalar coefficients $a$, $b$ and $c$ are computed through the vector inner products as $a = \left[\vector{G}_1\right]^{\top}_{_{3 \times 1}} \singlecontraction \left[\vector{G}_1\right]_{_{3 \times 1}}$, $b = \left[\vector{G}_1\right]^\top_{_{3 \times 1}} \singlecontraction \left[\vector{G}_2\right]_{_{3 \times 1}}$, and $c = \left[\vector{G}_2\right]^\top_{_{3 \times 1}} \singlecontraction \left[\vector{G}_2\right]_{_{3 \times 1}}$. 

\item The local orthonormal basis vectors $\left[(\vector{E}_1)_{i,j}\right]_{_{3 \times 1}}$ and $\left[(\vector{E}_2)_{i,j}\right]_{_{3 \times 1}}$
are computed using the covariant vectors $\left[(\vector{G}_1)_{i,j}\right]_{_{3 \times 1}}$ and $\left[(\vector{G}_2)_{i,j}\right]_{_{3 \times 1}}$, 
with $\left[(\vector{E}_1)_{i,j}\right]_{_{3 \times 1}}$ chosen parallel to $\left[(\vector{G}_1)_{i,j}\right]_{_{3 \times 1}}$, 
and then $\left[(\vector{E}_2)_{i,j}\right]_{_{3 \times 1}}$  is taken orthogonal to $\left[(\vector{E}_1)_{i,j}\right]_{_{3 \times 1}}$, i.e., 
\begin{equation}
\left[(\vector{E}_1)_{i,j}\right]_{_{3 \times 1}} = \left[(\vector{G}_1)_{i,j}\right]_{_{3 \times 1}},
\end{equation}
and 
\begin{equation}
\left[(\vector{E}_2)_{i,j}\right]_{_{3 \times 1}} = \frac{\left[(\vector{G}_2)_{i,j}\right]_{_{3 \times 1}} -\left(\left[\vector{E}_1\right]^\top_{_{3 \times 1}} \singlecontraction \left[\vector{G}_2\right]_{_{3 \times 1}} \right) \left[\vector{E}_1\right]_{_{3 \times 1}} }{ || \left[\vector{G}_2\right]_{_{3 \times 1}} - \left( \left[\vector{E}_1\right]^\top_{_{3 \times 1}} \singlecontraction  \left[\vector{G}_2\right]_{_{3 \times 1}} \right) \left[\vector{E}_1\right]_{_{3 \times 1}} || }.
\end{equation}

\item The deformed configuration tangent vectors $\left[(\vector{a}_1)_{i,j}\right]_{_{3 \times 1}}$ and $\left[(\vector{a}_2)_{i,j}\right]_{_{3 \times 1}}$ are obtained using the deformed position vectors, and the Euclidean distance between the position vectors in the reference configuration as 

\begin{equation}
\left[(\vector{a}_1)_{i,j}\right]_{_{3 \times 1}} = \frac{\left[\vector{x}_{i+1, j}\right]_{_{3 \times 1}} - \left[\vector{x}_{i-1, j}\right]_{_{3 \times 1}} }{ ||\left[\vector{X}_{i+1, j}\right]_{_{3 \times 1}} - \left[\vector{X}_{i-1, j}\right]_{_{3 \times 1}} || },
\label{vec_a1}
\end{equation}
and
\begin{equation}
\left[(\vector{a}_2)_{i,j}\right]_{_{3 \times 1}} = \frac{\left[\vector{x}_{i, j+1}\right]_{_{3 \times 1}} - \left[\vector{x}_{i, j-1}\right]_{_{3 \times 1}} }{ ||\left[\vector{X}_{i, j+1}\right]_{_{3 \times 1}} - \left[\vector{X}_{i, j-1}\right]_{_{3 \times 1}}|| }.
\label{vec_a2}
\end{equation}
The deformed unit normal vector $\left[(\vector{a}_3)_{i,j}\right]_{_{3 \times 1}}$ is obtained from the vector 
cross product between $\left[\vector{a}_1\right]_{_{3 \times 1}}$ and $\left[\vector{a}_2\right]_{_{3 \times 1}}$ as
\begin{equation}
\left[(\vector{a}_3)_{i,j}\right]_{_{3 \times 1}} = \frac{\left[(\vector{a}_1)_{i,j}\right]_{_{3 \times 1}} \times \left[(\vector{a}_2)_{i,j}\right]_{_{3 \times 1}}}{|| \left[(\vector{a}_1)_{i,j}\right]_{_{3 \times 1}} \times \left[(\vector{a}_2)_{i,j}\right]_{_{3 \times 1}} ||}.    
\end{equation}

\item The deformed configuration contravariant tangent vectors $\left[(\vector{a}^1)_{i,j}\right]_{_{3 \times 1}}$ and $\left[(\vector{a}^2)_{i,j} \right]_{_{3 \times 1}}$ are obtained using the reference configuration covariant tangent vectors $\left[(\vector{a}_1)_{i,j}\right]_{_{3 \times 1}}$ and $\left[(\vector{a}_2)_{i,j}\right]_{_{3 \times 1}}$ as follows
\begin{equation}
\left[(\vector{a}^1)_{i,j}\right]_{_{3 \times 1}} = \frac{1}{e^2 - df} \left( e \, \left[(\vector{a}_2)_{i,j}\right]_{_{3 \times 1}} - f \, \left[(\vector{a}_1)_{i,j}\right]_{_{3 \times 1}} \right),
\end{equation}
and
\begin{equation}
\left[(\vector{a}^2)_{i,j}\right]_{_{3 \times 1}} = \frac{1}{e^2 - df} \left( e \, \left[(\vector{a}_1)_{i,j}\right]_{_{3 \times 1}} - d \, \left[(\vector{a}_2)_{i,j}\right]_{_{3 \times 1}} \right),
\end{equation}
where the scalar coefficients $d$, $e$, and $f$ are given by the vector inner product as 
$d = \left[(\vector{a}_1)_{i,j}\right]^{\top}_{_{3 \times 1}} \singlecontraction \left[\vector{a}_1\right]_{_{3 \times 1}}$, $e = \left[\vector{a}_1\right]^{\top}_{_{3 \times 1}} \singlecontraction \left[(\vector{a}_2)_{i,j}\right]_{_{3 \times 1}}$, and $f = \left[(\vector{a}_2)_{i,j}\right]^{\top}_{_{3 \times 1}} \singlecontraction \left[(\vector{a}_2)_{i,j}\right]_{_{3 \times 1}}$.

\item The derivatives of the deformed configuration tangent vectors,   
$\left[ (\vector{a}_{1, 1})_{i,j} \right]_{_{3 \times 1}}$, $\left[(\vector{a}_{1, 2})_{i,j} \right]_{_{3 \times 1}}$,
$\left[(\vector{a}_{2, 1})_{i,j} \right]_{_{3 \times 1}}$, and $\left[(\vector{a}_{2, 2})_{i,j} \right]_{_{3 \times 1}}$,
with respect to curvilinear coordinates are computed as follows 
\begin{equation}
\left[ (\vector{a}_{1, 1})_{i,j} \right]_{_{3 \times 1}} = \frac{\left[\left(\vector{a}_1\right)_{i+1, j} \right]_{_{3 \times 1}} - \left[\left(\vector{a}_1\right)_{i-1, j}\right]_{_{3 \times 1}} }{ || \left[\vector{X}_{i+1, j}\right]_{_{3 \times 1}} - \left[\vector{X}_{i-1, j}\right]_{_{3 \times 1}} || },
\label{a_1,1}
\end{equation}
\begin{equation}
\left[(\vector{a}_{1, 2})_{i,j} \right]_{_{3 \times 1}} = \frac{\left[ \left(\vector{a}_1\right)_{i, j+1}\right]_{_{3 \times 1}} - \left[ \left(\vector{a}_1\right)_{i, j-1}\right]_{_{3 \times 1}} }{ ||\left[ \vector{X}_{i, j+1}\right]_{_{3 \times 1}} - \left[ \vector{X}_{i, j-1}\right]_{_{3 \times 1}}|| },
\label{a_1,2}
\end{equation}
\begin{equation}
\left[(\vector{a}_{2, 1})_{i,j} \right]_{_{3 \times 1}} = \frac{\left[ \left(\vector{a}_2\right)_{i+1, j}\right]_{_{3 \times 1}} - \left[ \left(\vector{a}_2\right)_{i-1, j}\right]_{_{3 \times 1}} }{ ||\left[ \vector{X}_{i+1, j}\right]_{_{3 \times 1}} - \left[ \vector{X}_{i-1, j}\right]_{_{3 \times 1}}|| },
\label{a_2,1}
\end{equation}
and 
\begin{equation}
\left[(\vector{a}_{2, 2})_{i,j} \right]_{_{3 \times 1}} = \frac{\left[ \left(\vector{a}_2\right)_{i, j+1}\right]_{_{3 \times 1}} - \left[ \left(\vector{a}_2\right)_{i, j-1}\right]_{_{3 \times 1}} }{ || \left[ \vector{X}_{i, j+1}\right]_{_{3 \times 1}} - \left[\vector{X}_{i, j-1}\right]_{_{3 \times 1}} || }\textbf{}.
\label{a_2,2}
\end{equation}

\item 
The first order shape function derivatives (defined in Equations \ref{matrix_dR_dtheta_alpha_components} and \ref{matrix_d2R_dtheta_alpha_beta_components}), 
and the second order shape function derivatives (defined in Equations \ref{dR_dt1} and \ref{dR_dt2}) are computed using the shape function matrix (Equation \ref{matrix_R_components}), as follows 
\begin{equation}
\left[\left(\frac{\partial R}{\partial \theta^{1}}\right)_{i, j} \right]_{_{3 \times 3N}} = \frac{\left[ R_{\, i+1, j} \right]_{_{3 \times 3N}} - \left[ R_{\, i-1, j} \right]_{_{3 \times 3N}}}{ || \left[\vector{X}_{i+1, j}\right]_{_{3 \times 1}} - \left[\vector{X}_{i-1, j}\right]_{_{3 \times 1}}||},
\label{dR_dt1}
\end{equation}

\begin{equation}
\left[\left(\frac{\partial R}{\partial \theta^{2}}\right)_{i, j} \right]_{_{3 \times 3N}} = \frac{\left[ R_{i, j+1} \right]_{_{3 \times 3N}} - \left[ R_{i, j-1} \right]_{_{3 \times 3N}} }{ || \left[\vector{X}_{i, j+1}\right]_{_{3 \times 1}} - \left[\vector{X}_{i, j-1}\right]_{_{3 \times 1}} ||},
\label{dR_dt2}
\end{equation}

\begin{align}
\left[ \left(\frac{\partial^2 R}{\partial \theta^{1} \partial \theta^{1} }\right)_{i, j} \right]_{_{3 \times 3N}} &= \frac{1}{ || \left[ \vector{X}_{i+1, j}\right]_{_{3 \times 1}} - \left[\vector{X}_{i-1, j}\right]_{_{3 \times 1}} ||} \left( \left[ \left( \frac{\partial R}{\partial \theta^{1}}\right)_{i+1, j} \right]_{_{3 \times 3N}} - \left[ \left( \frac{\partial R}{\partial \theta^{1}}\right)_{i-1, j} \right]_{_{3 \times 3N}} \right), \\
\left[ \left(\frac{\partial^2 R}{\partial \theta^{1} \partial \theta^{2} }\right)_{i, j} \right]_{_{3 \times 3N}} &= \frac{1}{ || \left[\vector{X}_{i+1, j}\right]_{_{3 \times 1}} - \left[\vector{X}_{i-1, j}\right]_{_{3 \times 1}} ||} \left( \left[ \left( \frac{\partial R}{\partial \theta^{2}}\right)_{i+1, j} \right]_{_{3 \times 3N}} - \left[ \left( \frac{\partial R}{\partial \theta^{2}}\right)_{i-1, j} \right]_{_{3 \times 3N}} \right), \\
\left[ \left(\frac{\partial^2 R}{\partial \theta^{2} \partial \theta^{1} }\right)_{i, j} \right]_{_{3 \times 3N}} &= \frac{1}{ || \left[\vector{X}_{i, j+1}\right]_{_{3 \times 1}} - \left[\vector{X}_{i, j-1}\right]_{_{3 \times 1}} ||} \left( \left[ \left( \frac{\partial R}{\partial \theta^{1}}\right)_{i, j+1} \right]_{_{3 \times 3N}} - \left[ \left(  \frac{\partial R}{\partial \theta^{1}}\right)_{i, j-1} \right]_{_{3 \times 3N}} \right), \\
\left[ \left(\frac{\partial^2 R}{\partial \theta^{2} \partial \theta^{2} }\right)_{i, j} \right]_{_{3 \times 3N}} &= \frac{1}{ || \left[ \vector{X}_{i, j+1}\right]_{_{3 \times 1}} - \left[\vector{X}_{i, j-1} \right]_{_{3 \times 1}}||} \left( \left[ \left( \frac{\partial R}{\partial \theta^{2}}\right)_{i, j+1} \right]_{_{3 \times 3N}} - \left[ \left(  \frac{\partial R}{\partial \theta^{2}}\right)_{i, j-1} \right]_{_{3 \times 3N}} \right).
\label{d2R_dtab}
\end{align}
Thus, we have all the quantities required to calculate the internal force vector $\left[ \vector{F}_{\text{int}} \right]$. In order to ensure efficient implementation, 
several quantities needed at the Gauss points, that remain constant during deformation, are computed only once and stored efficiently. 
\end{enumerate}

\subsubsection{Computations at the element edge} 
\label{boundary_edge_computations}

\begin{figure}[ht!]
     \centering
     \begin{subfigure}[b]{0.45\textwidth}
         \centering
         \includegraphics[scale=.5]{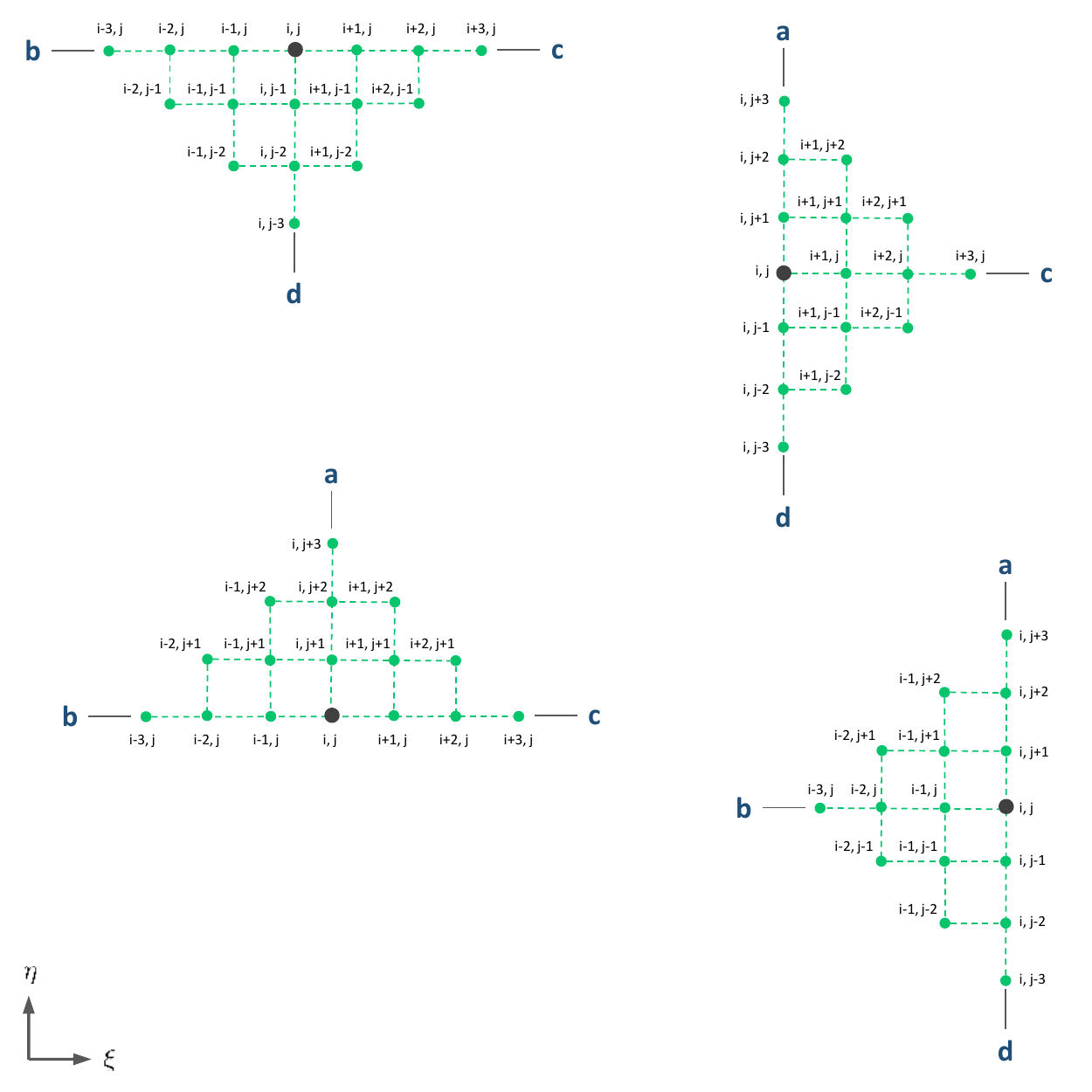}
         \caption{Case 1: $\eta=1$ on the edge `bc'.}
         \label{boundary_network_case_1}
     \end{subfigure}
     \hfill
     \begin{subfigure}[b]{0.45\textwidth}
         \centering
         \includegraphics[scale=1.0,trim={9cm 6.5cm 0cm 0cm},clip]{Images/Edge_network.pdf}
         \caption{Case 2: $\xi=0$ on the edge `ad'.}
         \label{boundary_network_case_2}
     \end{subfigure}
     \quad
     \begin{subfigure}[b]{0.45\textwidth}
         \centering
         \includegraphics[scale=1.0,trim={0cm 4cm 5cm 5cm},clip]{Images/Edge_network.pdf}
         \caption{Case 3: $\eta=0$ on the edge `bc'.}
         \label{boundary_network_case_3}
     \end{subfigure}
     \hfill
     \begin{subfigure}[b]{0.45\textwidth}
         \centering
         \includegraphics[scale=1.0,trim={8cm 0cm 0cm 6.25cm},clip]{Images/Edge_network.pdf}
         \caption{Case 4: $\xi=1$ on the edge `ad'.}
         \label{boundary_network_case_4}
     \end{subfigure}
\caption{Schematics for asymmetric half-networks at four different parametric edges.}
\label{edge_network_cases}
\end{figure}

Since the NURBS domain $\Xi_1\times \Xi_2$, discussed in Section ~\ref{sec:nurbs}, is mapped to the shell midsurface, the outer edge of the shell midsurface can be identified with 
four parametric edges: $\eta = 0$ and $\xi$ $\in$ $[0,1]$, $\eta = 1$ and $\xi$ $\in$ $[0,1]$, $\eta$ $\in$ $[0,1]$ and $\xi=0$, and  $\eta$ $\in$ $[0,1]$ and $\xi=1$, in the NURBS space. To compute the normal derivative of the shape functions $\left[\frac{\partial R}{\partial \nu} \right]_{_{3 \times 3N}}$, i.e. derivative with respect to the curvilinear 
coordinate $\nu$ which is normal to the edge, we cannot initialize a symmetric network on an edge, unlike inside the element domain (as discussed in the previous section). 
Thus, an asymmetric half-network is initialized on the four edges, corresponding to the four parametric edges in the NURBS space, shown in Figure ~\ref{edge_network_cases}. We use the second order forward or backward schemes to compute the derivatives at the edges. Corresponding to the four parametric edges, the normal derivatives of the shape functions are computed as follows.

\textit{Case 1}: 
\begin{align}
\left[ \left( \frac{\partial R}{\partial \theta^{2}} \right)_{i, j} \right]_{_{3 \times 3N}} &= \dfrac{-3 \left[ R_{i, j-2}\right]_{_{3 \times 3N}} + 4 \left[ R_{i, j-1} \right]_{_{3 \times 3N}} - \left[ R_{i, j} \right]_{_{3 \times 3N}}}{ 2 || \left[ \vector{X}_{i, j} \right]_{_{3 \times 1}} - \left[ \vector{X}_{i, j-1} \right]_{_{3 \times 1}}||},
\label{case_1_R,1_and_R,2}
\end{align}

\textit{Case 2}: 
\begin{align}
\left[ \left(  \frac{\partial R}{\partial \theta^{1}} \right)_{i, j} \right]_{_{3 \times 3N}} &= \dfrac{ -3 \left[ R_{i, j}\right]_{_{3 \times 3N}} + 4 \left[ R_{i+1, j}\right]_{_{3 \times 3N}} - \left[ R_{i+2, j} \right]_{_{3 \times 3N}}}{ 2 || \left[ \vector{X}_{i+1, j} \right]_{_{3 \times 1}} - \left[ \vector{X}_{i, j} \right]_{_{3 \times 1}}||},
\label{case_2_R,1_and_R,2}
\end{align}

\textit{Case 3}: 
\begin{align}
\left[ \left( \frac{\partial R}{\partial \theta^{2}} \right)_{i, j} \right]_{_{3 \times 3N}} &= \dfrac{ -3 \left[ R_{i, j}\right]_{_{3 \times 3N}} + 4 \left[ R_{i, j+1}\right]_{_{3 \times 3N}} - \left[ R_{i, j+2}\right]_{_{3 \times 3N}}}{ 2 || \left[ \vector{X}_{i, j+1}\right]_{_{3 \times 1}} - \left[ \vector{X}_{i, j}\right]_{_{3 \times 1}}||},
\label{case_3_R,1_and_R,2}
\end{align}

\textit{Case 4}: 
\begin{align}
\left[ \left(  \frac{\partial R}{\partial \theta^{1}} \right)_{i, j} \right]_{_{3 \times 3N}} &= \dfrac{-3 \left[ R_{i-2, j} \right]_{_{3 \times 3N}} + 4 \left[ R_{i-1, j} \right]_{_{3 \times 3N}} - \left[ R_{i, j} \right]_{_{3 \times 3N}}}{ 2 || \left[ \vector{X}_{i-1, j}\right]_{_{3 \times 1}} - \left[ \vector{X}_{i, j}\right]_{_{3 \times 1}}||}.
\label{case_4_R,1_and_R,2}
\end{align}

\section{Dynamic Relaxation Method}
\label{sec:dynamic-relaxation}
Dynamic Relaxation (DR) is a numerical method which has been used effectively to 
obtain the static equilibrium solutions of a mechanical system, by creating a psuedo dynamical system with damping, and allowing
it to relax (by dissipating the kinetic energy) to arrive at an equilibrium state
\cite{metzger2003adaptive,rombouts2018equivalence,joldes2011adaptive,underwood1986dynamic,barnes1999form,brew1971non}.
Technicalities aside, the DR method exploits Lyapunov stability, in the sense that a dynamical system for which a Lyapunov function exists, once released from an initial state, 
can converge to a stable equilibria. The advantage of using dynamic relaxation is that, unlike iterative Newton-Raphson method, as used in conventional finite element, there is no requirement of constructing global stiffness matrices (which can become signular in presence of bifurcation). DR can be based on an explicit time integration scheme,
where by using appropriate mass scaling the critical size of the time increment can be increased. It should be noted that for a nonlinear system the convergence of DR may depend on 
the choice of initial state, and in cases nonlinear elastic shells more than one stable equilibria can exist for a given set of boundary conditions. Another advantage of using 
DR is that we do not need to seed the reference structure with artificial ``imperfections'' to trigger a particular deformation mode; however, to obtain different equilibrium configurations 
we will need to use different initial states. 

 In what follows, 
$ \Big[ \vector{${{u}}$} \Big]$, $ \Big[ \vector{${\dot{u}}$} \Big]$, $ \Big[ \vector{${\ddot{u}}$} \Big]$, $ \Big[ \vector{F}_{ext} \Big]$, and $\Big[ \vector{F}_{int} \Big]$ denote the global vectors, of size $3N\times1$ (we drop the size subscripts for sake of convenience), for displacements, velocities, accelerations, external forces, and internal forces, respectively, corresponding to different degrees of freedom for each control point.  The internal force vector $\Big[ \vector{F}_{int} \Big]$ includes contributions from 
$\left[ \vector{F}_{\text{s}} \right]$, $\left[ \vector{F}_{\text{G}} \right]$, $\left[ \vector{F}_{\text{p}} \right]$, and $\left[ \vector{F}_{\text{sp}} \right]$ (discussed in Section ~\ref{sec:FEA-discretization}). $\Big[ \tso{M}\Big]$ denotes the diagonal mass matrix, and $C$ is the damping coefficient. 
The system of equations for the finite element mesh is then given as

\begin{equation}
\Big[ \tso{M} \Big] \singlecontraction \Big[ \vector{$\ddot{u}$} \Big]+ C \Big[ \vector{$\dot{u}$} \Big] + \Big[ \vector{F}_{int} \Big] = \Big[ \vector{F}_{ext}\Big].
\label{DR_equation}
\end{equation}

We use a central difference scheme in time to solve Equation ~\ref{DR_equation} numerically. If $n$ denotes the discrete time step then computation of acceleration, the updates for velocity and displacement vectors are give as
 
\begin{equation}
\Big[ \vector{$\ddot{u}$} \Big]_{n} = \tso{M}^{-1} \singlecontraction \left(  \Big[  \vector{F}_{ext} \Big]_{n} - \Big[ \vector{F}_{int} \Big]_{n} - C \Big[ \vector{$\dot{u}$} \Big]_{n - \frac{1}{2}} \right),
\label{dr_eq_1}
\end{equation}

\begin{equation}
\Big[ \vector{$\dot{u}$} \Big]_{n + \frac{1}{2}} = \Big[ \vector{$\dot{u}$} \Big]_{n - \frac{1}{2}} + \Delta t \Big[ \vector{$\ddot{u}$} \Big]_{n},
\label{dr_eq_2}
\end{equation}

\begin{equation}
\Big[ \vector{u} \Big]_{n + 1} = \Big[ \vector{u} \Big]_{n} + \Delta t \Big[ \vector{$\dot{u}$} \Big]_{n + \frac{1}{2}},
\label{dr_eq_3}
\end{equation}
where,  $n + \frac{1}{2}$ is the midpoint between $n$ and $n+1$ time steps, and  $\Delta t = t_{n+1}-t_{n}$ . The mechanical system is perturbed from the reference state, and allowed to relax such that over time its kinetic energy is damped out. 
 A constant diagonal mass $\tso{M}$ is used, 
with $\tso{M} = m \tso{I}$, where $m$ is the constant nodal mass, 
and $\tso{I}$ is the second order identity tensor.
We note that time increments, damping coefficient, and nodal masses are simply fictous parameters, however, there selection 
must be done correctly to ensure numerical stability and convergence. 

Although there are several variations of DR method, such as adaptive DR (with dynamic parameters, e.g., \cite{joldes2011adaptive}), which can converge faster to the equilibrium solution, 
we have used a standard  DR with only three algorithmic parameters 
nodal mass ($m$), damping coefficient ($c$), and time step size ($\Delta t$). 
Kinetic energy of the total system was used as the termination criterion, and during iterations whenever the 
kinetic energy of the system went below the pre-specified threshold ($\epsilon_t$)  the algorithm was terminated.
An efficient, parallel, and scalable implementation for DR algorithm was carried out. 
A pseudo code for the implemented algorithm is presented in Algorithm \ref{dr_pseudo_code}.\\\\

\section{High-Peformance Computing-based DR}
\label{parallel_computations}
The current generation of CPUs offer multicore architecture in which a single physical processor comprises the core logic of multiple processors. The commercially available CPUs typically 
contain  8 to 64 cores, where each core can additionally provide two threads to increase the utilization of a single core. We define one physical machine as a node that contains multicore 
CPUs, RAM, hard disk, and other hardware peripherals. In this work, we achieve parallelism by utilizing multiple nodes, with each node containing multiple cores and multiple threads.
In \cite{ModernFortran}, Milan implemented a simple server and client program written in Fortran programming language using C language bindings and socket programming, where 
messages were exchanged between server and client programs running on separate processes. The messages were converted into bytes, and  sent over the network. We followed a similar 
strategy. Streaming the data directly over the network minimizes the transfer overheads compared to alternate approaches such as file reading and writing, especially over thousands of DR 
iterations. We benchmarked the server and client program, and found it suitable for our requirements. We also extended the newly developed server and client program to pass double-precision vectors of variable lengths from one server to multiple clients connected over a high-bandwidth intranet network.

\begin{figure}[hbt!]
    \centering
    \includegraphics[scale=0.5, trim={1.5cm 4cm, 2cm 0},clip]{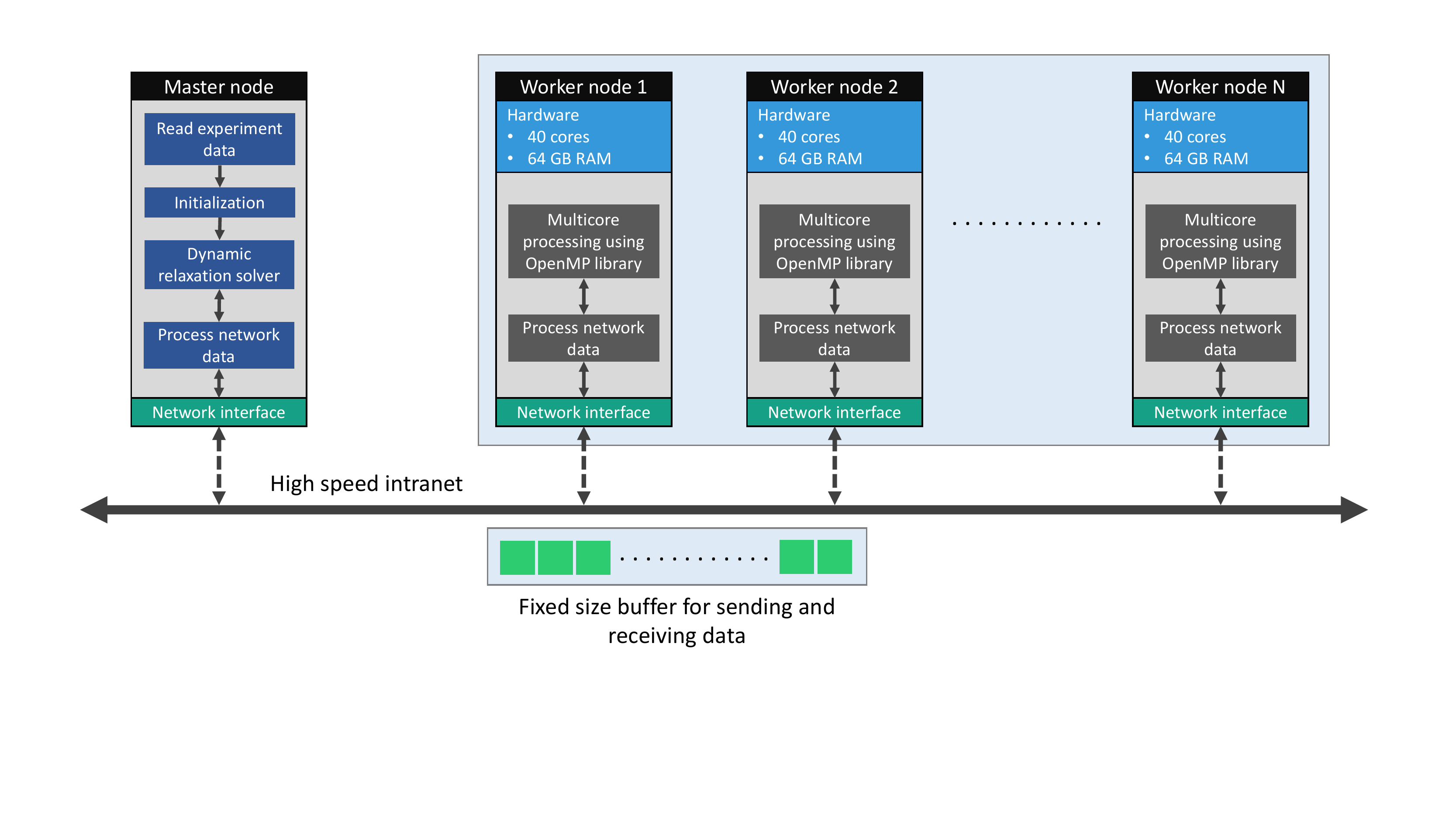}
    \caption{A schematic for the master and worker nodes for high-performance computing-based Dynamic Relaxation method developed in this study.}
    \label{parallel_setup}
\end{figure}

High-performance computing in a cluster environment leverages multiple nodes connected with a high-speed intranet network. A scheduler allocates requested resources to the queued 
jobs submitted by the users, and starts executing them on a priority basis. We utilize such a cluster to set up a master-slave work environment. A master node runs the dynamic relaxation 
algorithm, and sends the data to multiple slave nodes. A slave is any worker node containing multiple cores (up to 32) with two threads per core. Multi-threaded parallel computing is 
achieved using the OpenMP library on each worker node.
Figure \ref{parallel_setup} demonstrates the parallel environment with one master, and multiple worker nodes. The internal force vector is computed in parallel for each dynamic relaxation 
iteration. We divide the total number of elements ($P$) into the $N$ equal size chunks containing the unique elements. The role of a worker node is to receive one chunk of elements along 
with the current displacement vector, and return the internal force vector of the contributing chunk of elements, along with their bending and membrane energies. Each chunk gets allocated to 
one worker node, and all the chunks are evaluated in parallel. By utilizing such parallelism, we can evaluate $P/N$ elements on $N$ nodes faster than the serial setup (where all elements 
are evaluated sequentially).

A vector of size 3$N$ + 2 is sent to each worker node where $N$ is the total number of control points. $3N$ represents the size of displacement vectors. The remaining two slots indicate  the element indices sent to a particular slave node for processing. 
Once the worker nodes have computed the contributions to the internal force vector, and other quantities (like energies), the results are sent back to the master node in a vector of size $3N + 4T$, where $3N$ corresponds to the size of internal force vector, and $4T$ corresponds to the element numbers, bending energies, membrane energies, and strain energies of the elements. The data returned from the worker nodes are later processed by the master node.
Overall we were able to gain a speed-up proportional to number of worker nodes. We noted, however, that in a heterogeneous cluster environment, where worker nodes had different hardware specifications (e.g., different CPU frequencies, different RAM speeds), the slowest node was the bottle neck, and dominated the computational time required to complete one dynamic relaxation iteration. Secondly, even if one node failed to send data back to the master node, then the algorithm failed since the results from all the worker nodes are required to proceed to the next iteration. Using excessive worker nodes (beyond 10) also introduced noticeable data transfer overheads. We are working currently to overcome some of these limitations. \\\\

\begin{breakablealgorithm}
\caption{Pseudo code for the method of dynamic relaxation}
\label{dr_pseudo_code}
\begin{algorithmic}[1]
\Procedure{Dynamic Relaxation}{}

\State \emph{Initialization}:
\State $\Big[\vector{$\dot{u}$}\Big]_{-\frac{1}{2}}$ $\gets$ $ \vector{0}$ 
\State $\Big[\vector{u}\Big]_{0}$ $\gets$ $ \vector{0}$ 
\State $\Big[\vector{$\dot{u}$}\Big]_{0}$ $\gets$ $ \vector{0}$ 
\State Read parameters m, c and $\Delta t$ \Comment{File read}

\State \emph{Loading and relaxation iterations}:
\State count = 1 \Comment{Global iteration counter}
\For{( i = 1 to L ) } \Comment{L = Number of loading steps}

    \State Set $\vector{F}_{ext}$ for loading step \Comment{Apply load}
    \State Read $\vector{u}^{\text{given}}$ \Comment{Given displacements for current step}

    \For{( j = 1 to N ) } \Comment{N = Iterations per loading step}
    
        \State $\Big[ \vector{F}_{int} \Big]_n \gets \Big[\vector{u}\Big]_{n}$ \Comment{Compute internal force}
        
        \State $\Big[\vector{$\ddot{u}$}\Big]_{n} \gets m, c, \Big[ \vector{F}_{int} \Big]_n, \Big[ \vector{F}_{ext} \Big]_n,  \Big[ \vector{$\dot{u}$} \Big]_{n - \frac{1}{2}}$ \Comment{Use equation \ref{dr_eq_1}}
        
        \State $\Big[ \vector{$\dot{u}$} \Big]_{n + \frac{1}{2}} \gets \Delta t, \Big[ \vector{$\dot{u}$} \Big]_{n - \frac{1}{2}}, \Big[ \vector{$\ddot{u}$} \Big]_{n}$ \Comment{Use equation \ref{dr_eq_2}}
        
        \State $\Big[ \vector{u} \Big]_{n + 1} \gets \Delta t, \Big[ \vector{u} \Big]_{n}, \Big[ \vector{$\dot{u}$} \Big]_{n + \frac{1}{2}}$ \Comment{Use equation \ref{dr_eq_3}}

        \State $\Big[ \vector{u} \Big]_{n + 1} \gets \vector{u}^{\text{given}}$ \Comment{Set known displacements}
        
        \If{mod(count, K) == 0} \Comment{Save solution every K iterations}
            \State Save $\Big[ \vector{u} \Big]_{n + 1}$ 
        \EndIf
    
        \If{i == L}
            \State K.E. $\gets$ m, $\Big[\vector{$\dot{u}$}\Big]_{n+\frac{1}{2}}$ \Comment{Compute kinetic energy of system}
            \State Check termination  \Comment{For the last loading step}
        \EndIf

        \State $\Big[ \vector{u} \Big]_{n}$ $\gets$ $\Big[ \vector{u} \Big]_{n + 1}$ \Comment{Iterative update for displacement}
        \State $\Big[ \vector{$\dot{u}$} \Big]_{n-\frac{1}{2}}$ $\gets$ $\Big[ \vector{$\dot{u}$} \Big]_{n + \frac{1}{2}}$ \Comment{Iterative update for velocity}
        
        \State count $\gets$ count + 1 \Comment{Increment iteration counter}
    
    \EndFor
    
    \State Save $\Big[ \vector{u} \Big]_{n + 1}$ \Comment{One loading step is complete}

\EndFor

\State Compute element energies
\State Generate solution files

\EndProcedure
\end{algorithmic}
\end{breakablealgorithm}

\subsection{Guidelines for parameter selection}
The selection of DR parameters ($m$, $c$, and $\Delta t$) is done so as to satisy the CFL condition,
and therefore for a given problem the choice of these parameters depends on the geometry of the finite element mesh, and material properties. For a
particular SVK material with $\lambda$,  and $\mu$ as the material parameters, and 
$h_e$ as the minimum finite element size, we first 
estimate the the critical density ($\rho_{cr}$), based on the dilataional wave speed, and chosen
time increment $\Delta t$ as
\begin{equation}
\rho_{cr} = \left(\frac{\Delta t}{h_e} \right)^2 \times (\lambda + 2 \mu).
\label{rho_cr}
\end{equation}
We note that for a given  $\Delta t$, if the chosen material density is greater than $\rho_{cr}$, then the time increment ($\Delta t$)
will satisfy the CFL condition.  
Using this $\rho_{cr}$, in conjunction with reference shell 
surface area ($A$) and thickness ($t$), we then calculate the critical mass ($m_{cr}$) as
\begin{equation}
m_{cr} = \rho_{cr} \times A \times t.
\label{m_cr}
\end{equation}
The  $m_{cr}$ is distributed equally on each control point (where $N_{cp}$ denotes the total number of control points), and is scaled using a  factor $\rho_s$ as
\begin{equation}
m = \frac{m_{cr}}{N_{cp}} \times \rho_s.
\label{DR_mass_expression}
\end{equation}

The damping parameter $c$ is chosen in the range from $10^{-1}$ to $10^{-4}$. We defer the rigorous numerical analysis of DR method for nonlinear problems for the future. 
\section{Numerical Results}
\label{sec:numerical-results}
Typically in wrinkling  or buckling (instability) problems, in commercial softwares like Abaqus, first a linear analysis is carried 
out to identify the eigen-modes of the structure. Then, an imperfection, in an ad hoc manner, repsented by the scaled combinations of 
the obtained eigen modes, is 
seeded into the desired structure, and then the Riks analysis is carried out. When subject to loading, the structure deforms in 
compliance with the seeded imperfection, and continues to deform 
into the nonlinear or postbucking regime. The final equilibrium state is thus 
dependent on the choice of the eigen modes, and the scale factors used while seeding the imperfection. In contrast, in 
simulations with DR, we do not seed any imperfection in the structure, rather we perturb the structure from its initial configuration (which in fact is physically justified), 
and allow the structure to relax and reach an equilibrium state.

\subsection{Lifting and twisting of an annular sheet}

\begin{figure}[ht]
    \centering
    \includegraphics[scale=0.6]{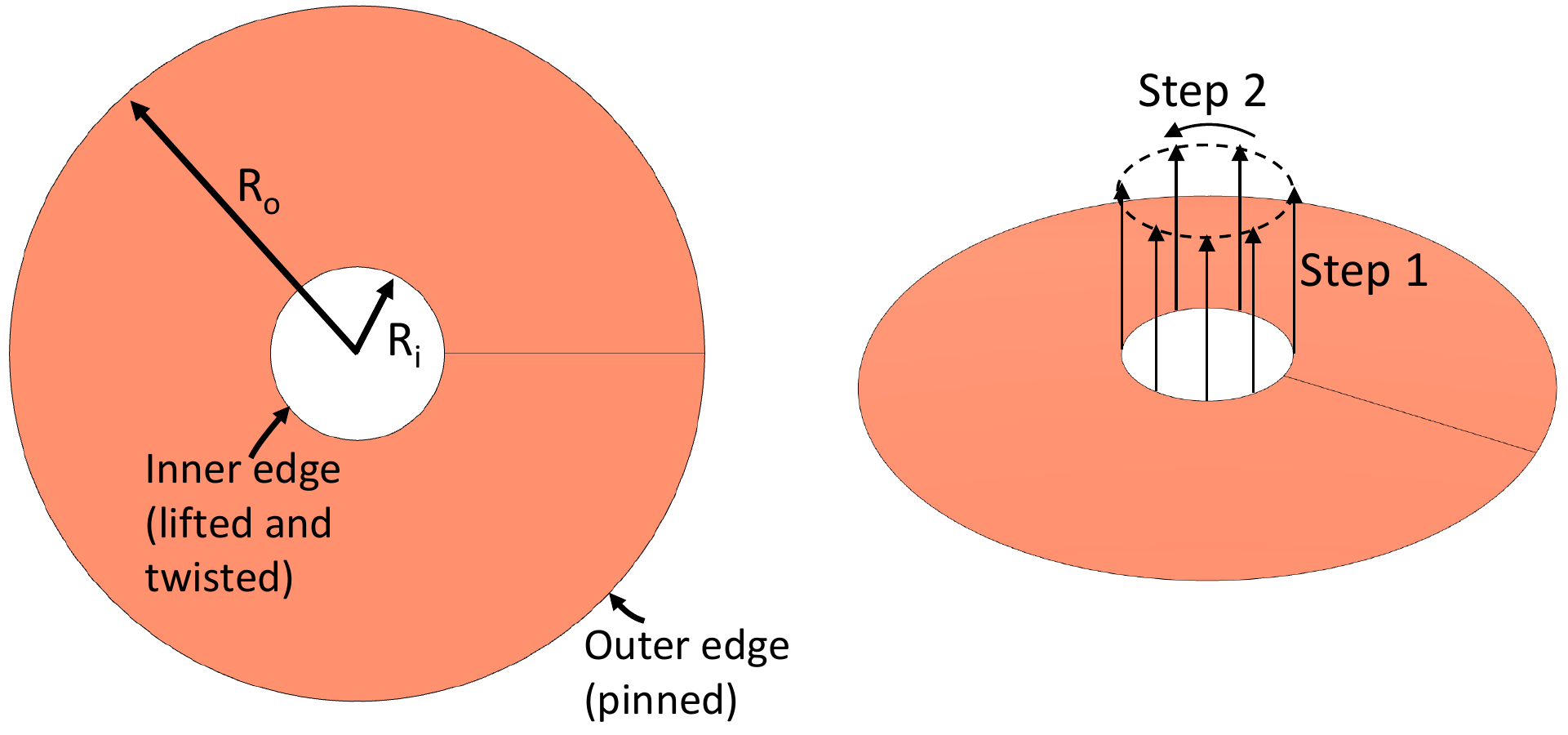}
    \caption{A schematic for the reference configuration of the annular sheet subject to lifting and twisting.}
    \label{ref-annular-sheet}
\end{figure}

In this example (taken from \cite{steigmann2013koiter,haseganu1994analysis}), a flat annular sheet (shown in Figure ~\ref{ref-annular-sheet}) is pinned on the outer edge, and the inner edge is lifted and twisted. This mode of deformation ultimately leads to generation of wrinkles. 
The inner and outer radii of the sheet are set to $62.5$ mm and $250$ mm, respectively. A total of 2,560 NURBS 
elements (10 in radial direction and 256 in the circumferential direction), of degree 3, with 
a total number of 3,354 control points were used. 
The sheet thickness is taken to be 0.025 mm, and elastic modulus, and Poisson's were chosen as 
3,500 tonne mm$^{-1}$ sec$^{-2}$, and 0.31, respectively.
The outer boundary is pinned (i.e. displacement is restricted but rotation is allowed), and the inner edge is displaced vertically 
out  of the plane (lifting step) by 125 mm, followed by a $90^\circ$ anticlockwise rotation (twisting step). There is stretching dominated deformation in the lifting step, and during twisting, as will be seen, the sheet relaxes into wrinkled 
patterns by accommodating fine-scale bending.

\begin{figure}[ht!]
    \centering
    \includegraphics[scale=0.5]{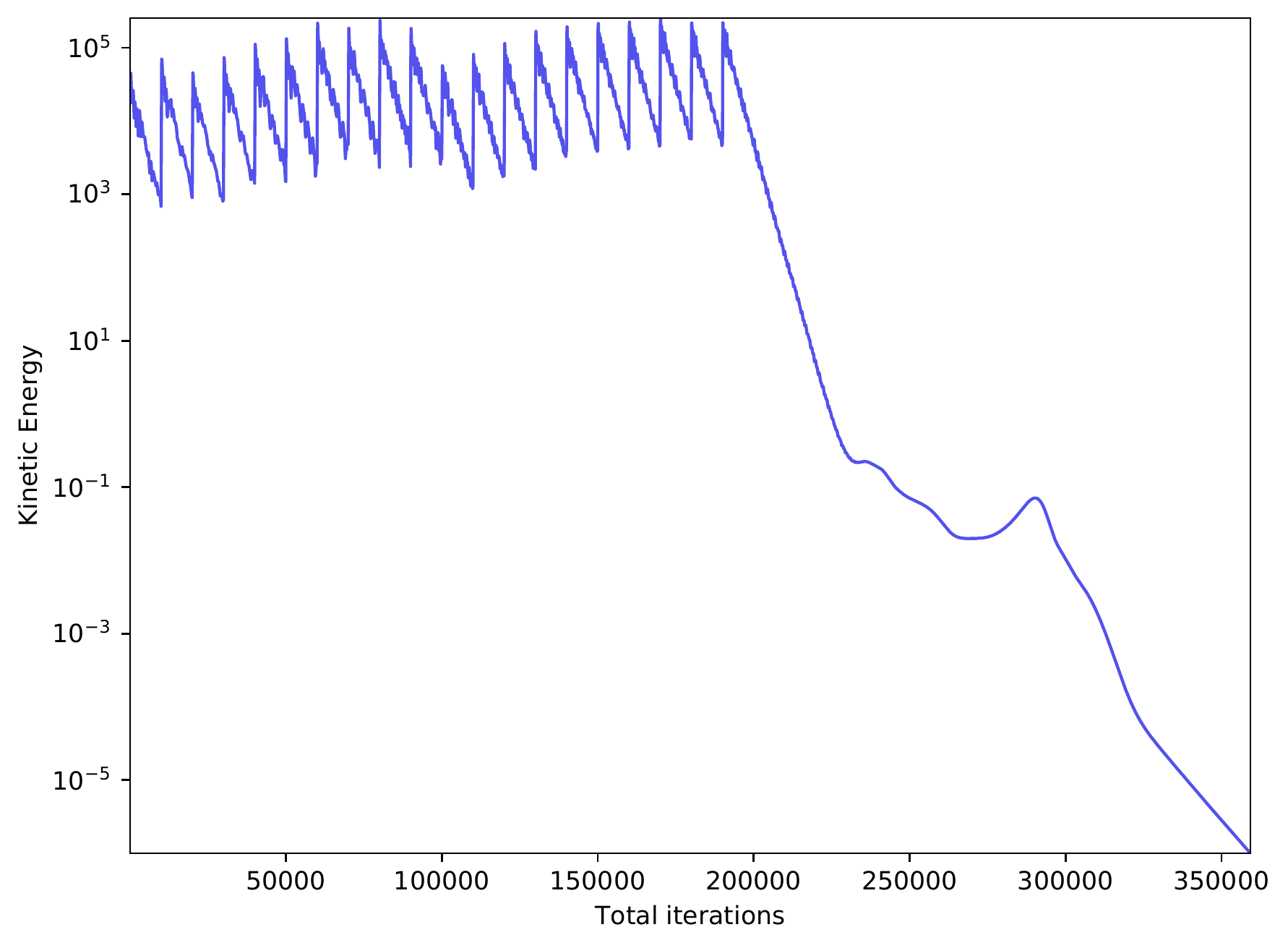}
    \caption{The progress of dynamic relaxation algorithm for the pull-out and twisting of an annular sheet}
    \label{Lift_rotate_DR_KE_Progress}
\end{figure}

\begin{figure}[h!]
\centering
\includegraphics[scale=0.4]{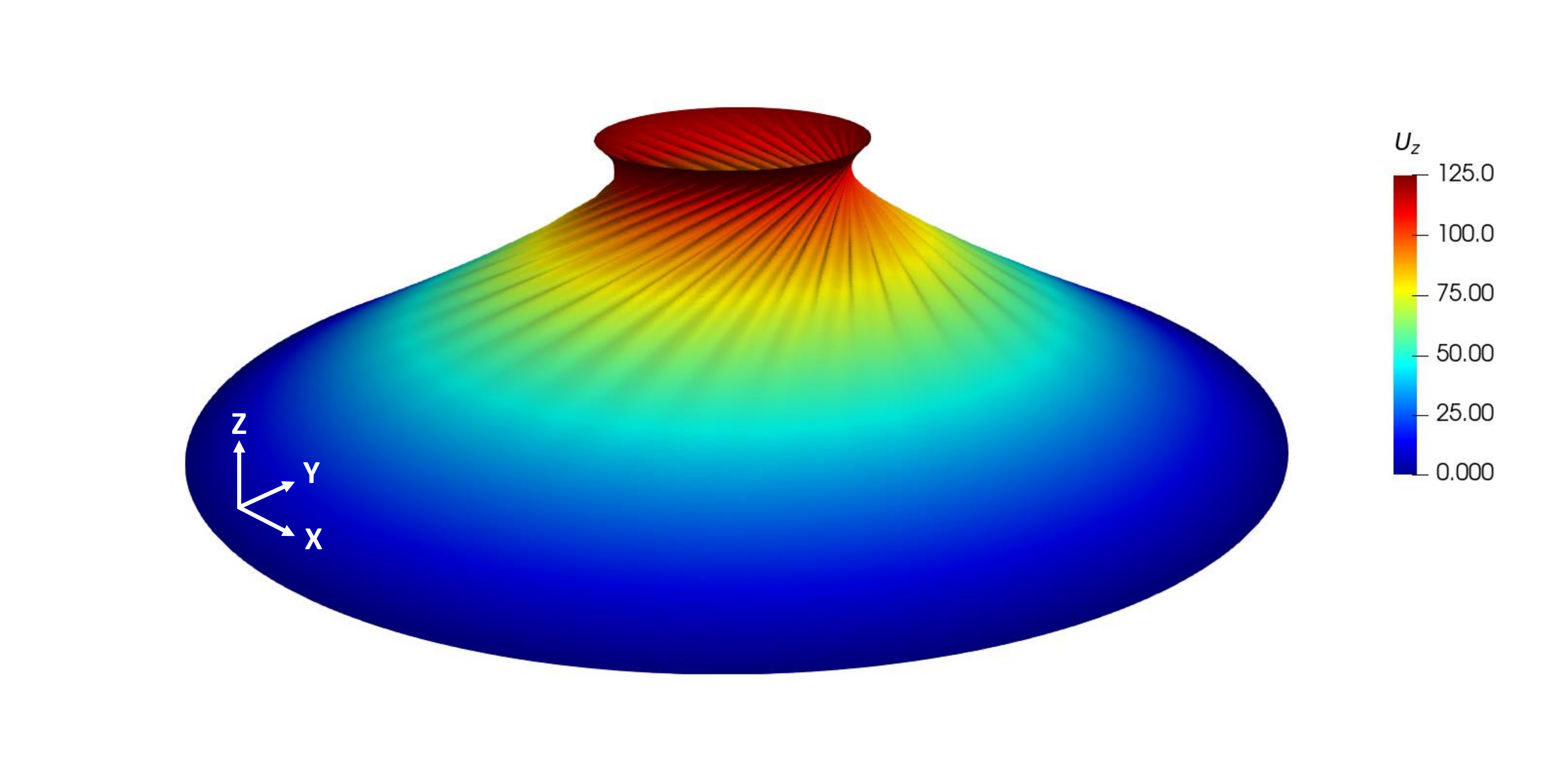} 
\caption{Obtained equilibrium solution for lifting and twisting of an annular sheet. The contours are shown for the z-displacement.}
\label{def_solution_LR}
\end{figure}

We estimated DR parameters as $m$ = $0.282$, $c$ = $0.1$, and choose $\Delta t$ = $0.001$ for this problem. 
Both, lifting and twisting steps, were applied in 10 increments each.  Kinetic energy 
termination criterion of $\epsilon_t=10^{-6}$ was used in the last loading increment, whereas in every other prior incremental loading the DR was run for 10,000 iterations. Figure 
\ref{Lift_rotate_DR_KE_Progress} shows the evolution of kinetic energy with respect to total DR iterations across all increments.
The obtained equilibrium solution is shown in Figure \ref{def_solution_LR}. Equilibrium 
bending, and membrane energies were found to be $1.549$ and $4.095 \times 10^{6}$, respectively. 
It should be noted that equilibrium membrane energy is orders of magnitude larger than the bending energy; however, 
without accounting for bending effects in the shell model, wrinkles cannot be generated.

\begin{figure}[h!]
\centering
\includegraphics[scale=0.4]{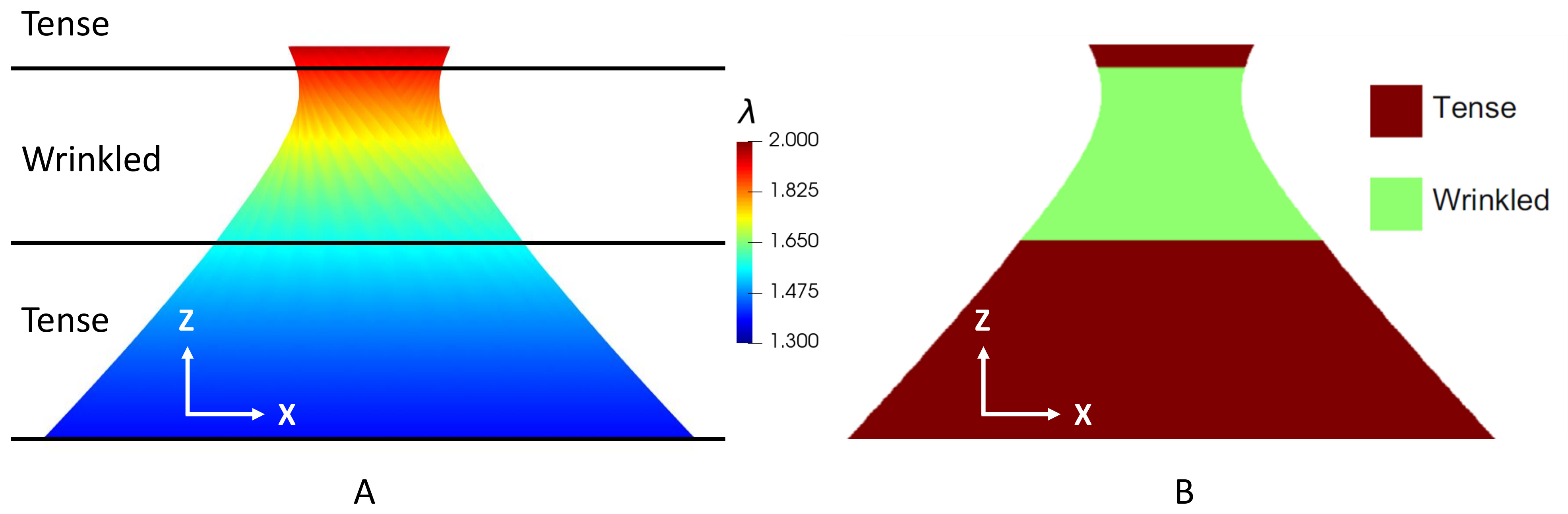}
\caption{Comparison of ``Tense'' and ``Wrinkled'' regions obtained using (A) our NURBS-based formulation, and (B) tension-field theory membrane model obtained from \cite{steigmann2013koiter}.}
\label{LR_region_compare}
\end{figure}

\begin{figure}[h!]
\centering
\includegraphics[scale=0.45]{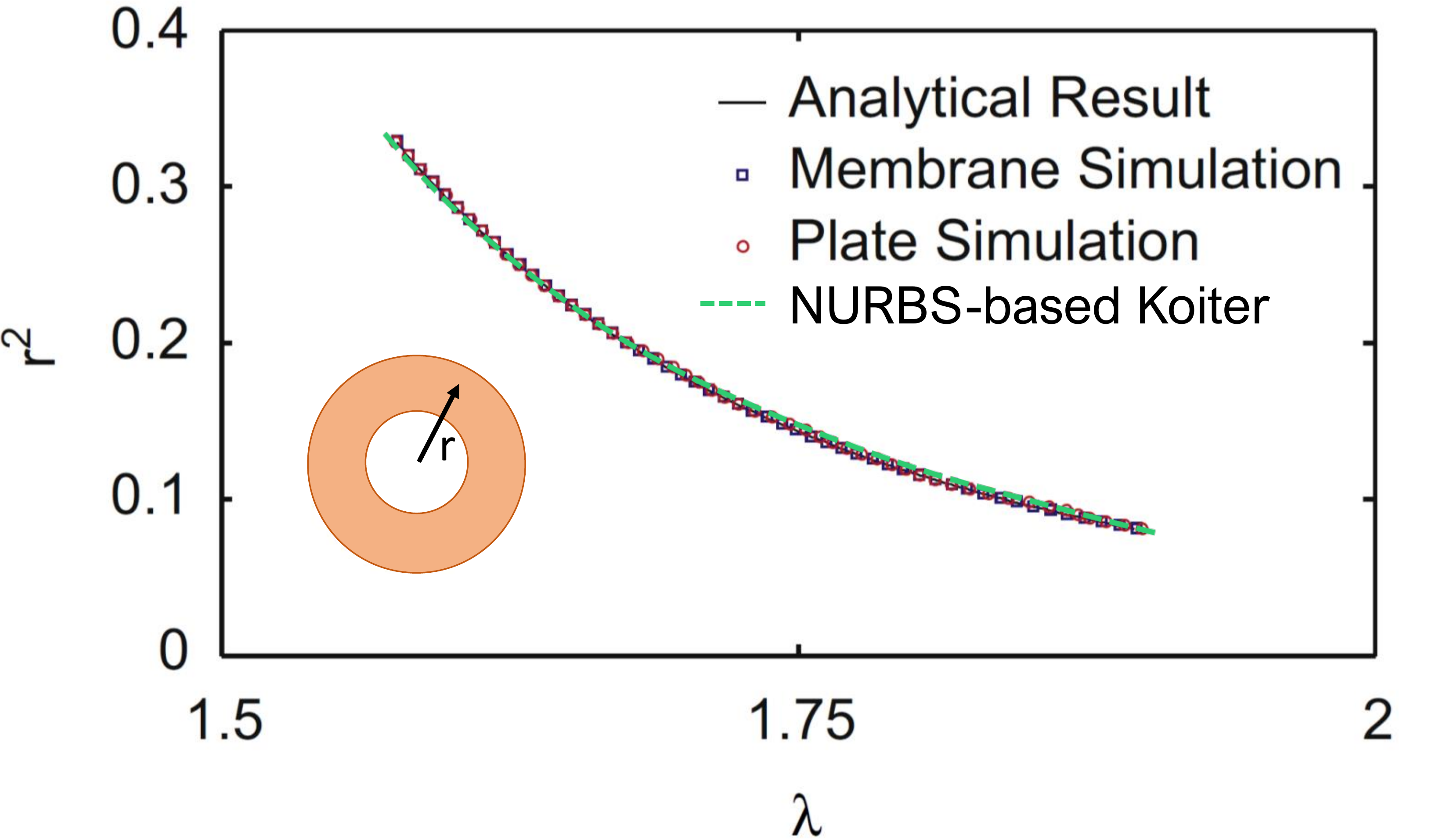}
\caption{Variation of maximum principal stretch ($\lambda$) with $(r/R_o)^2$ for the wrinkled region.}
\label{LR_r2_lambda_compare}
\end{figure}

\begin{figure}[h!]
\centering
\includegraphics[scale=0.4]{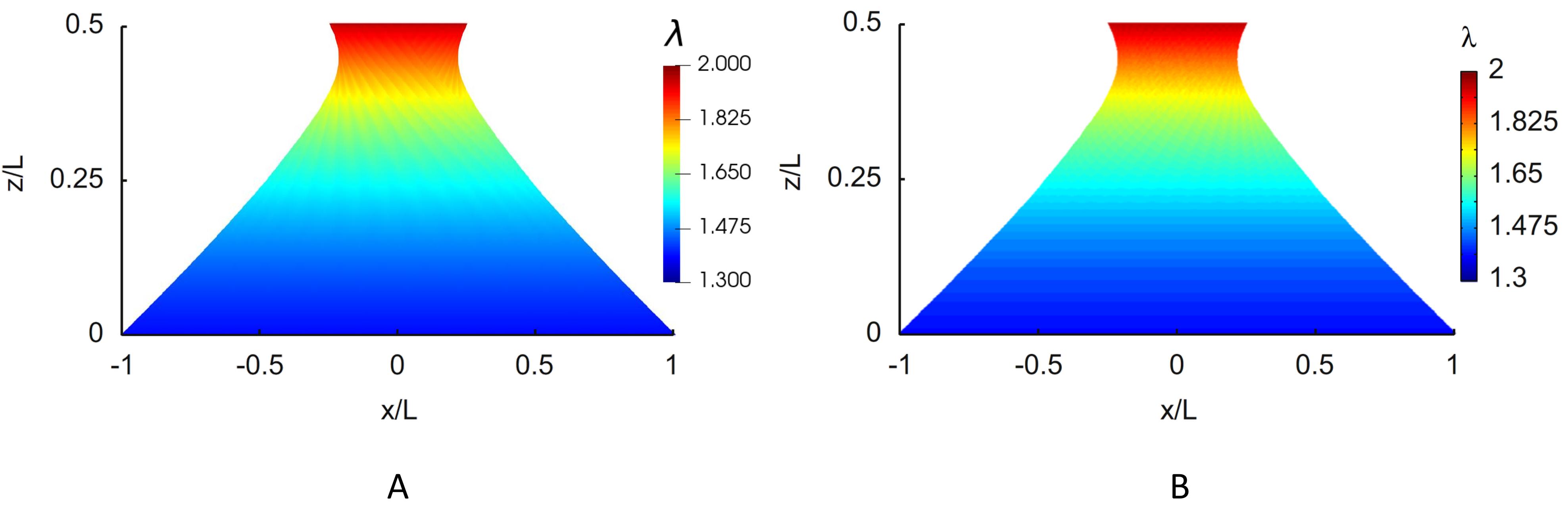}
\caption{Comparison of maximum principal stretch ($\lambda$) contours of the equilibrium configurations obtained according to (A) our simulations, and (B)  reported in \cite{steigmann2013koiter}.}
\label{LR_lambda_comparion}
\end{figure}

In Figures ~\ref{LR_region_compare}A-B, we compare our results with those presented in 
\cite{steigmann2013koiter}. Figure \ref{LR_region_compare}A shows the projected, two dimensional, view 
of the equilibrium solution from our simulations, and Figure \ref{LR_region_compare}B shows the equilibrium solution from \cite{steigmann2013koiter}. Like in \cite{steigmann2013koiter}, we are also able to find a wrinkled region 
between two ``Tense'' regions, and that the wrinkles are formed 
away from the top and bottom edges in the neck region. For quantitative comparisons, 
the maximum principal stretch ($\lambda$) in the wrinkled region for our equilibrium solution, 
tension field theory \cite{haseganu1994analysis}, and numerical simulations in \cite{steigmann2013koiter} are compared in
Figure ~\ref{LR_r2_lambda_compare}. These plots reveal an excellent match. 
Finally, the comparison of maximum principal stretch contours on the whole deformed domain
from our solution, and those obtained in \cite{steigmann2013koiter} are shown in 
~\ref{LR_lambda_comparion}A-B. These results once again exhibit an excellent agreement.

\clearpage

\subsection{Shearing of a rectangular sheet} \label{shear_sheet_reference}
\begin{figure}[ht]
    \centering
    \includegraphics[scale=0.5]{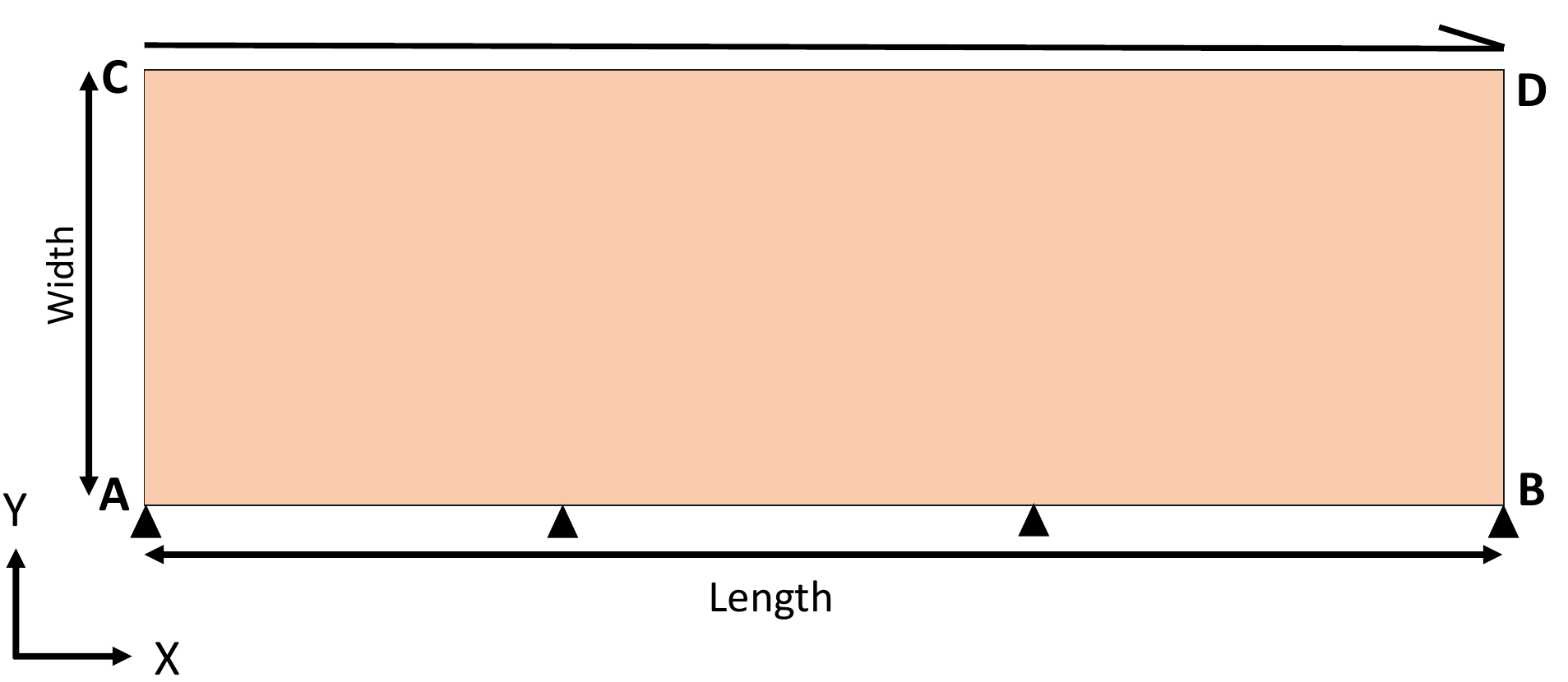}
    \caption{A schematic of a rectangular sheet for the shearing of Kapton sheet.}
    \label{ref-shear-sheet}
\end{figure}

The problem of shearing a rectangular Kapton sheet, and growth of wrinkles, has been studied computationally, and experimentally 
in \cite{wong2006wrinkled,steigmann2013koiter}. Figure \ref{ref-shear-sheet} shows a schematic of a planar 
rectangular sheet, with length, width, and thickness as 380 mm, 128 mm, and 0.025 mm, respectively. The sheet is modeled with 12,800 NURBS elements of degree 3 (with 320 and 40 elements the along X and Y directions, respectively). 
Young's modulus and Poisson's ratio are chosen as 3,500 tonne mm$^{-1}$ sec$^{-2}$ and 0.31, respectively.
A horizontal displacement of 0.5 mm is applied on the edge CD in the X direction, and the lower edge AB is pinned. 
The total shearing displacement is applied in 20 loading increments, the intermediate loading increments are run for a fixed number of 2,500 iterations, and the last increment utilizes a kinetic energy termination criterion of $\epsilon_t=10^{-6}$. 
The DR parameters  $m$, $c$, and $\Delta t$ are chosen as 0.0301, 0.01, and 0.001. 

\begin{figure}[ht!]
    \centering
    \includegraphics[scale=0.5]{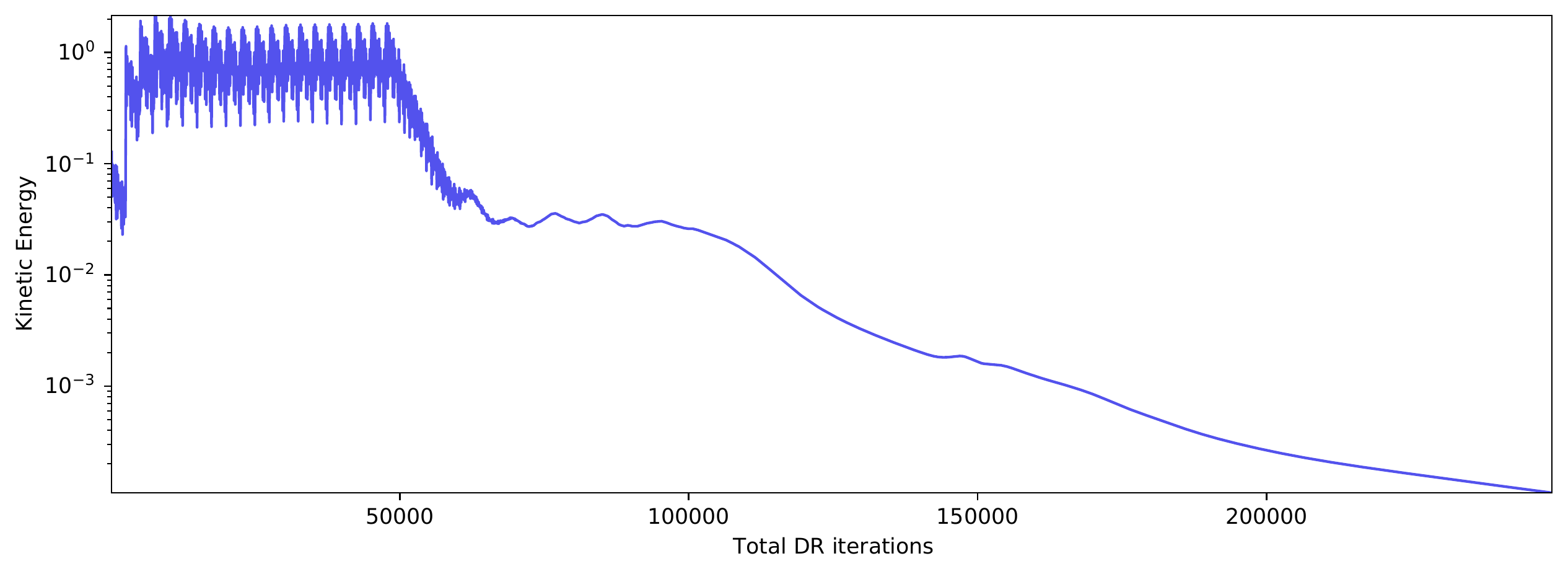}
    \caption{Kinetic energy of finite element shearing plate example with respect to dynamic relaxation iterations.}
    \label{Shear_KE_Progress}
\end{figure}

\begin{figure}[h!]
\centering
\includegraphics[scale=0.45]{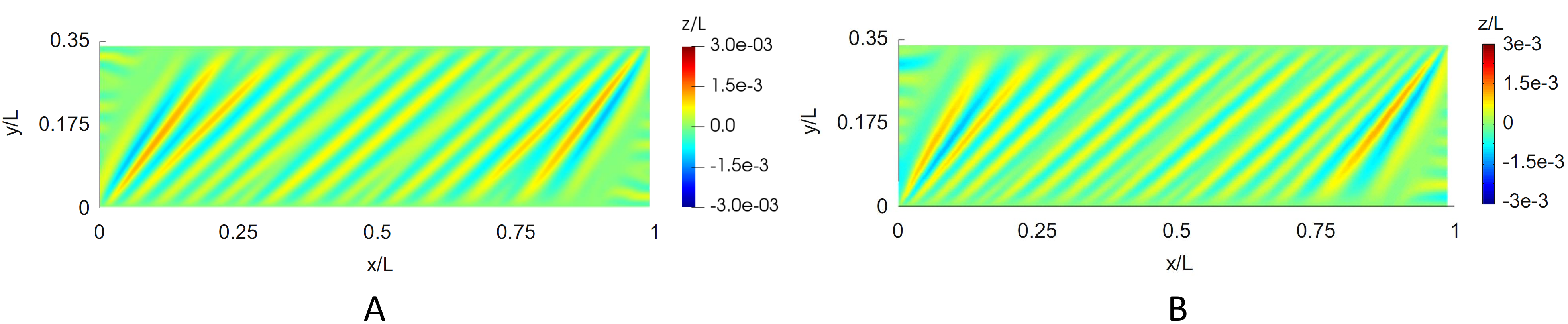} 
\caption{The equilibrium solutions for shearing of a rectangular Kapton sheet (A) NURBS-based Koiter, and (B) reported by \cite{steigmann2013koiter}.}
\label{def_solution_shear}
\end{figure}

The evolution of kinetic energy with respect to total DR iterations from all increments is shown in Figure \ref{Shear_KE_Progress}.
Figures \ref{def_solution_shear} A-B show the contour plots of the Z-displacements for the final equilibrium solutions obtained using our formulation, and the solution reported in \cite{steigmann2013koiter}. We observe that the obtained wrinkles from our formulation in Figure \ref{def_solution_shear} A are similar to those reported in Figure \ref{def_solution_shear} B. The wrinkles from these simulations are oriented at an angle of $45^\circ$ with respect to the lower edge, in the middle region of the sheet, and are also similar to those reported experimentally \cite{wong2006wrinkled}, shown in Figure \ref{def_solution_shear}.

\begin{figure}[h!]
\centering
\includegraphics[scale=0.65]{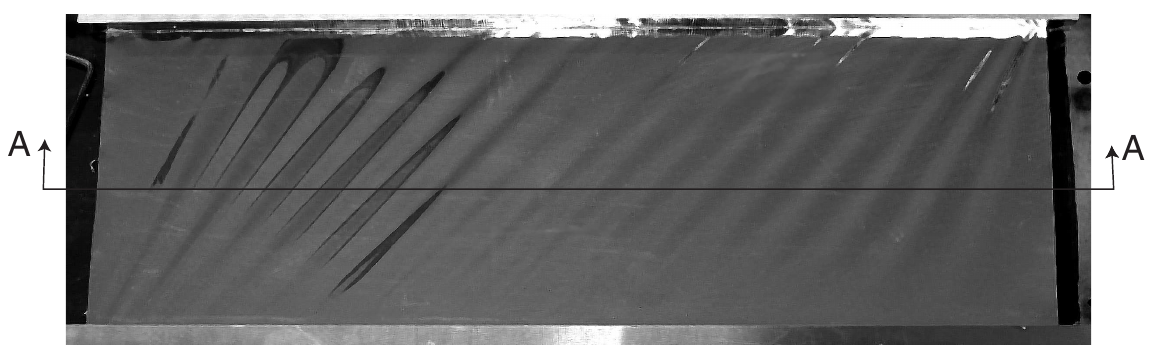} 
\caption{The experimental result for shearing of a rectangular Kapton sheet reported in \cite{wong2006wrinkled}.}
\label{experiment_solution_shear}
\end{figure}

\begin{figure}[h]
\centering
\includegraphics[scale=0.4]{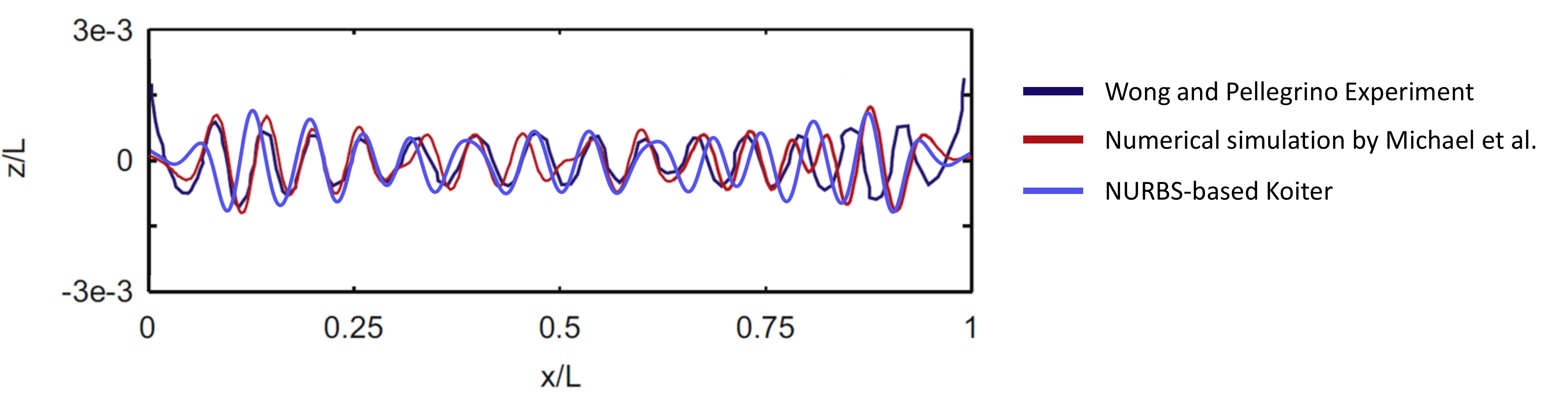} 
\caption{Comparison of wrinkle profiles obtained from experiments \cite{wong2006wrinkled}, reference numerical solution \cite{steigmann2013koiter}, and our formulation.}
\label{shear_profile_compare}
\end{figure}

We also compared the cross-sectional profiles of simulated wrinkles along a cross-section `AA' (shown in Figure ~\ref{experiment_solution_shear}), chosen at the half-width location,  with the experimental results from \cite{wong2006wrinkled}, and the numerical solution from \cite{steigmann2013koiter}. These comparisons are shown in Figure \ref{shear_profile_compare}. We note that our numerical simulation exhibits an excellent match with the reference numerical solution, and the experimental data, in the middle region of the sheet. 
At the boundaries, our solution and the reference numerical solution deviate from the reported experimental results. 
We suspect that these deviations could be experimental artificats due to material defects at the edges in the specimens, e.g., if a rectangular sheet of desired dimension is cut from a larger stock, then edges may undergo local plastic deformation,  
and bear residual stress states, which could lead to edge displacement profiles that cannot be modeled by just using the nonlinear elastic formulation as done in our simulations, or the reference numerical solutions. 

\subsection{Point indentation of a semi-cylinder}
\begin{figure}[h!]
\centering
\includegraphics[scale=0.5]{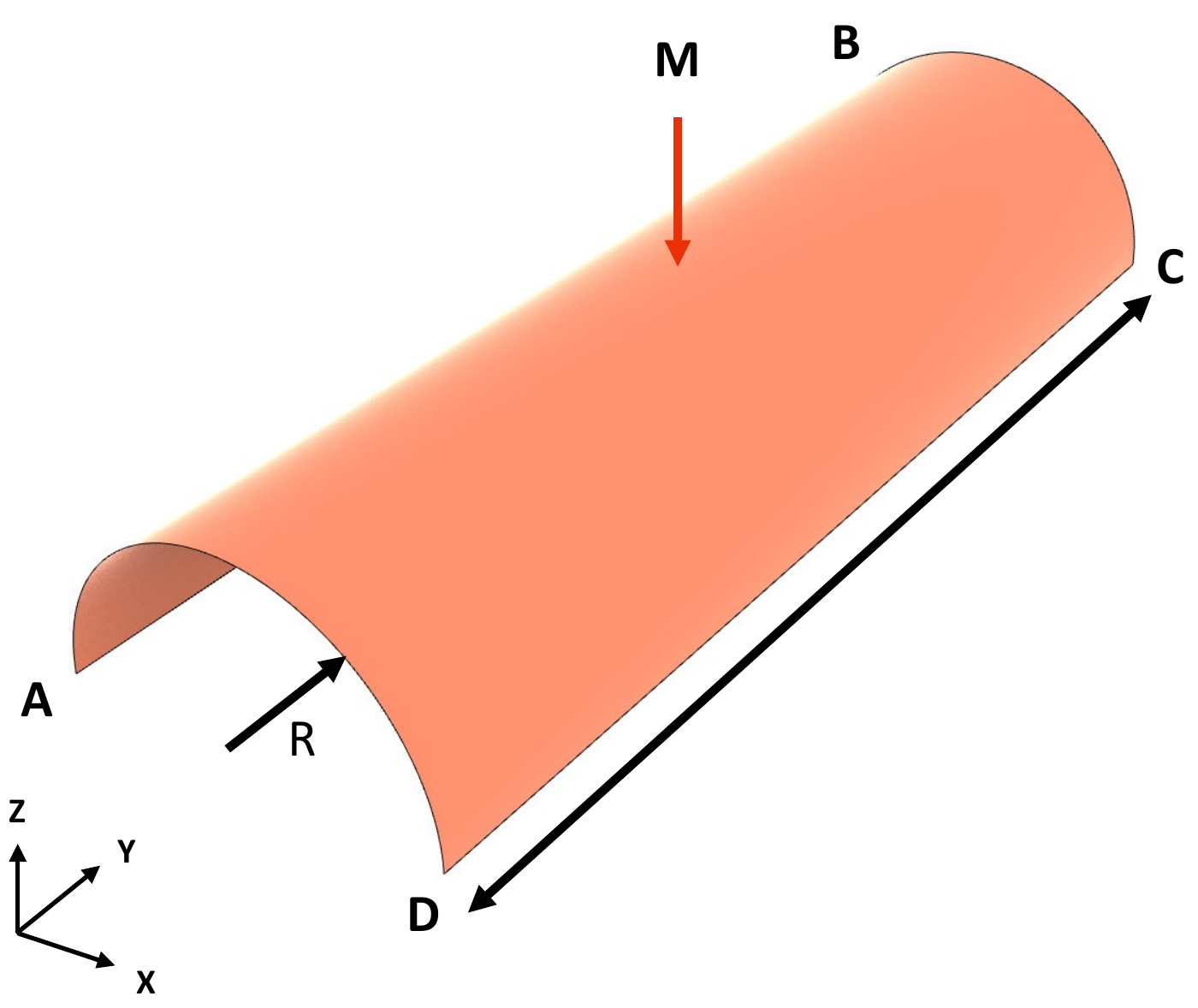} 
\caption{Indentation of a semi-cylinder under point load.}
\label{MSC_reference}
\end{figure}

In this experiment, we investigate the bifurcation of a semi-cylinder under point load as studied in \cite{vaziri2008localized}. Figure 
\ref{MSC_reference} shows the reference configuration of a semi-cylinder which is subject to point displacement. The edges AB and CD are clamped, and 
edges AD and BC are traction free. The length of CD is 6.0 mm, and the cylinder radius (R) is 1.0 mm. The thickness of cylinder is taken to be 0.015 
mm, with Young's modulus, and Poisson's ratio as 1000.0 tonne mm$^{-1}$ sec$^{-2}$, and 0.30, respectively. A displacement of 0.8 mm is applied 
in the negative Z-direction at $M$ in 20 loading steps.   A total of 512 NURBS elements (32 along the length and 16 along the circumferential direction), of degree 3, with 665 control points with rational 
weights were chosen. The DR parameters $m$, $c$, and $\Delta t$ were chosen as $0.0163$,  $0.01$, and $0.001$, respectively. A termination 
criterion of $\epsilon_t=10^{-14}$ was used in each loading step to obtain quasi-static intermediate equilibrium states, and accurate reaction force to describe the bifurcation behavior. The point displacement $M$, and clamped conditions at AB and CD were both enforced through penalty formulations (as discussed in Section ~\ref{sec:penalty-formulation}), using a penalty factor of 100. 
We also used the commercial FEA software Abaqus to model the point indentation of the semi-cylinder, with 5,200 S4R elements consisting of 5,353 nodes. Static general loading with adaptive stabilization factor of 0.05, and dissipated energy fraction of 0.0002 was used to tackle bifurcation. 
These were chosen following the simulation guidelines from \cite{vaziri2008localized}. NLGEOM analysis option was selected to accommodate for  geometric nonlinearity. 

\begin{figure}[h!]
\centering
\includegraphics[scale=0.4]{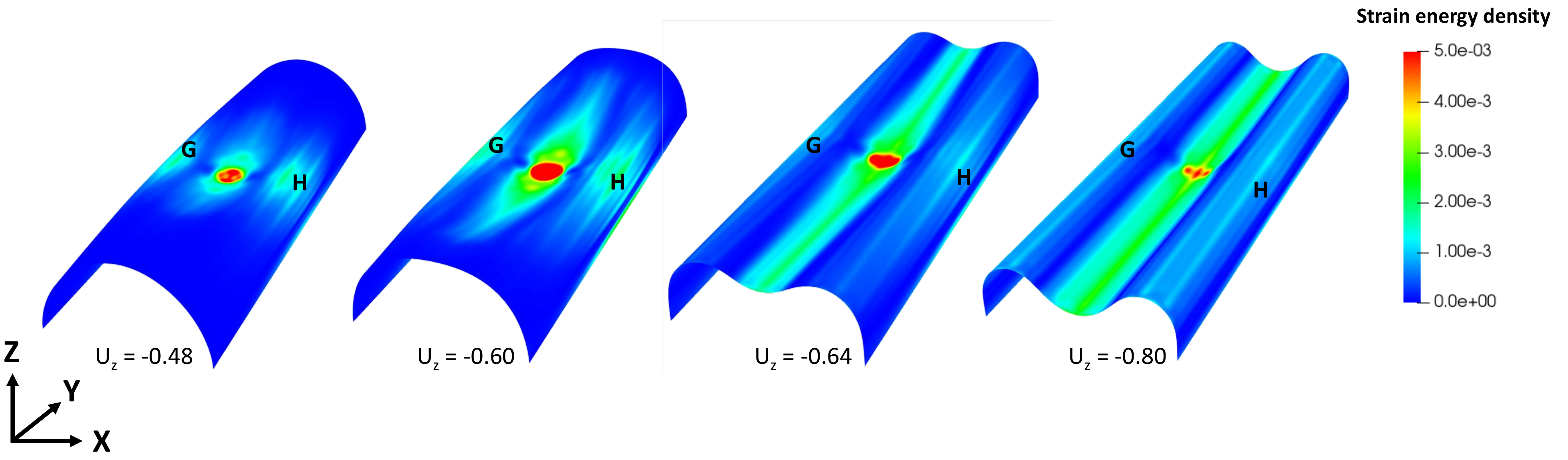} 
\caption{Evolution of the deformed configurations of the semi-cylinder under a point load. The contours represent strain energy density per unit area (bending and stretching).}
\label{msc_seds}
\end{figure}

The evolution of deformation under point indentation is shown in Figure ~\ref{msc_seds}. In the reference state, the cylinder that has zero Gaussian  curvature, and as the indentation proceeds first the deformation is localized in the region close to point displacement,
and then appearance of two new vertices, marked as G and H, occurs where the strain energy density begins to concentrate. Given that the 
cylinder has zero principal curvature along the length direction (Y-axis), and a negative curvature in the direction perpendicular to the length 
(X-axis), the kinematic compatibility of the shell surface under the point indentation causes a reduction in the negative curvature, and development of 
positive curvature along the length direction. The points G and H mark the locations of vertices formed, where positive curvature (generated due to point indentation) meets the negative curvature on the undeformed part of the cylindrical wall. Upon continued loading, 
local bending is accompanied by stretching, and eventually the deformation field extends and reaches the curved traction-free  boundaries, and the 
cylinder bifurcates into configuration with length-wise valley in the central region, while developing a positive curvature in the valley. The strain energy density 
remains highly concentrated under the point load, with a bending dominated core surrounded by a stretched region. 
\begin{table}[ht!]
\centering
\caption{Simulation data for the point indentation of a semi-cylinder in Abaqus, and our proposed formulation.}
\label{MSC_compare_params_with_abaqus}
\begin{tabular}{c | c c c c}
\hline
					& Number of elements  & Nodes      & Control points              & Strain energy 			\\              \hline
Abaqus S4R                 &   5,200    			&  5,353     & -                                  & $1.259 \times 10^{-2}$ \\
NURBS-based Koiter     &   512       			&  -            &  665				       & $1.433 \times 10^{-2}$ \\        
\hline
\end{tabular}
\end{table}

\begin{figure}[h!]
\centering
\includegraphics[scale=0.45]{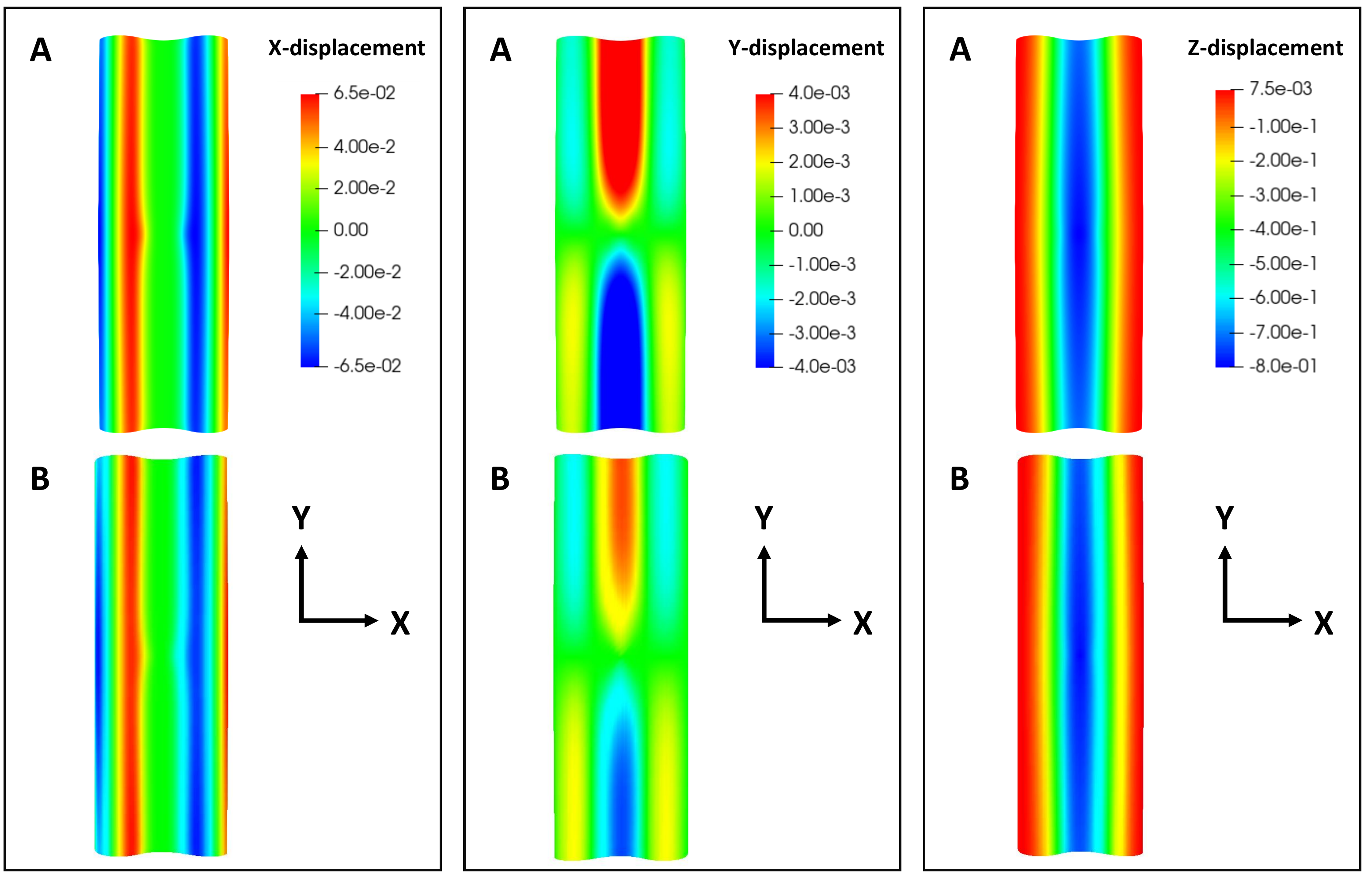} 
\caption{Comparison of obtained displacements from (A) NURBS-based Koiter and (B) Abaqus S4R.}
\label{msc_displacements}
\end{figure}
\begin{figure}[h!]
\centering
\includegraphics[scale=0.45]{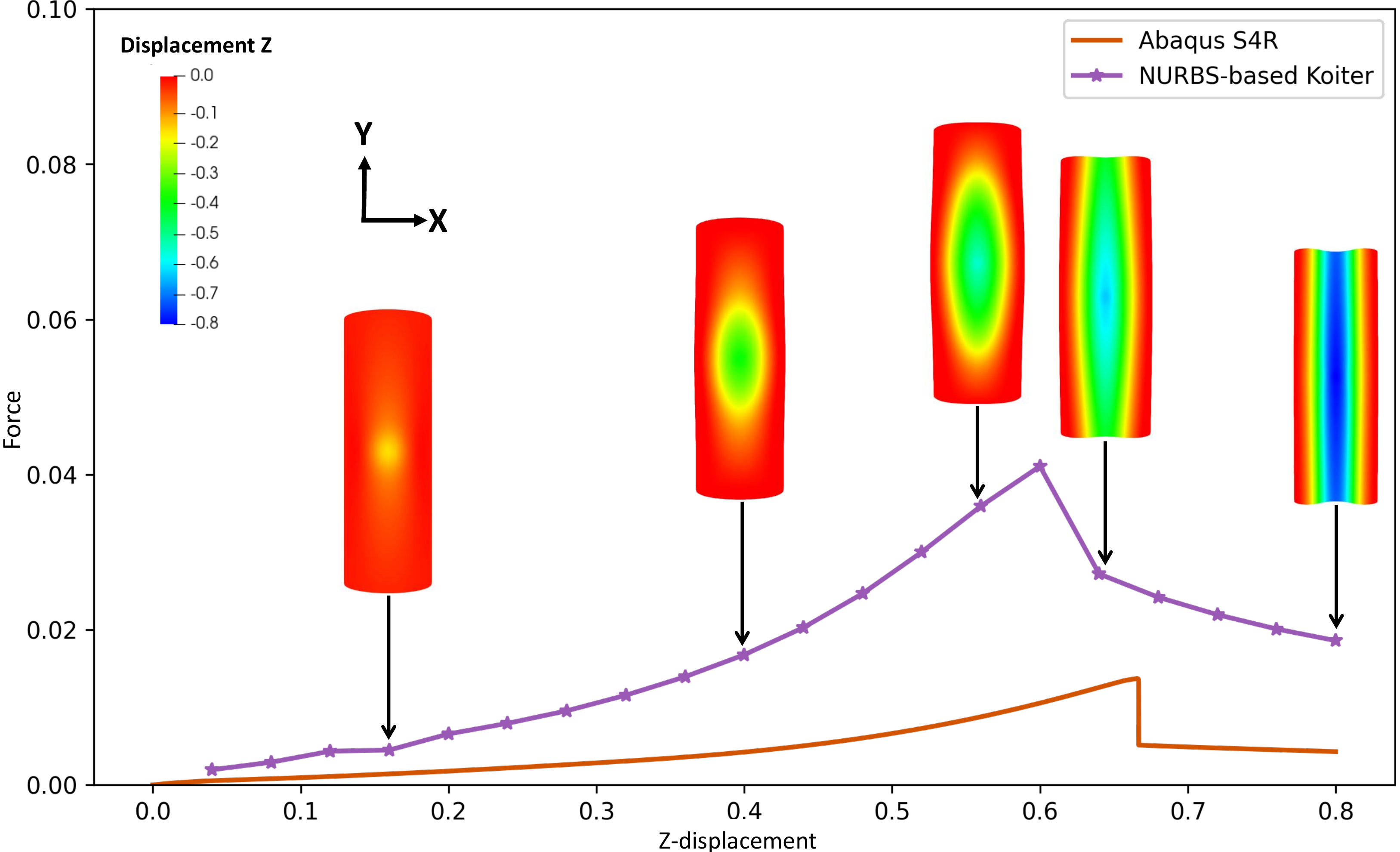} 
\caption{Force vs displacement curves for NURBS-based Koiter model, and Abaqus S4R.}
\label{MSC_F_vs_u}
\end{figure}

Figure \ref{msc_displacements} compares the final equilibrium displacement fields obtained from our formulation, and those from Abaqus. Clearly, there is a good match between the contours of all the displacement fields. Table \ref{MSC_compare_params_with_abaqus} 
summarizes the simulation parameters, and lists the total elastic strain energy of the cylinder at the final equilibrium step from Abaqus, and our formulation. 
We also compared the force vs displacement curves (at the location of the point indentation) from Abaqus, and our proposed formulation in Figure 
~\ref{MSC_F_vs_u}. Abaqus yields a lower force vs displacement curve. Upon further investigation, we noted that Abaqus S4R shell 
elements use Discrete Kirchhoff (DK) constraint, i.e., satisfaction of the Kirchhoff constraint at discrete points on the shell surface, and do not ensure 
tangent-continuity (in the weak sense) throughout the shell midsurface (and that the approach is in fact non-conforming, where jumps in normal 
derivatives are controlled by the imposed DK constraints), thus allowing developments of ``kinks''. On the other hand, NURBS-based formulation, 
by construction, maintains tangent-continuity throughout the computational domain during deformation. We refer to \cite{bernadoua1994convergence} for numerical analysis of finite elements based on DK constraints in the linear case.

\begin{figure}[t]
\centering
\includegraphics[scale=0.5]{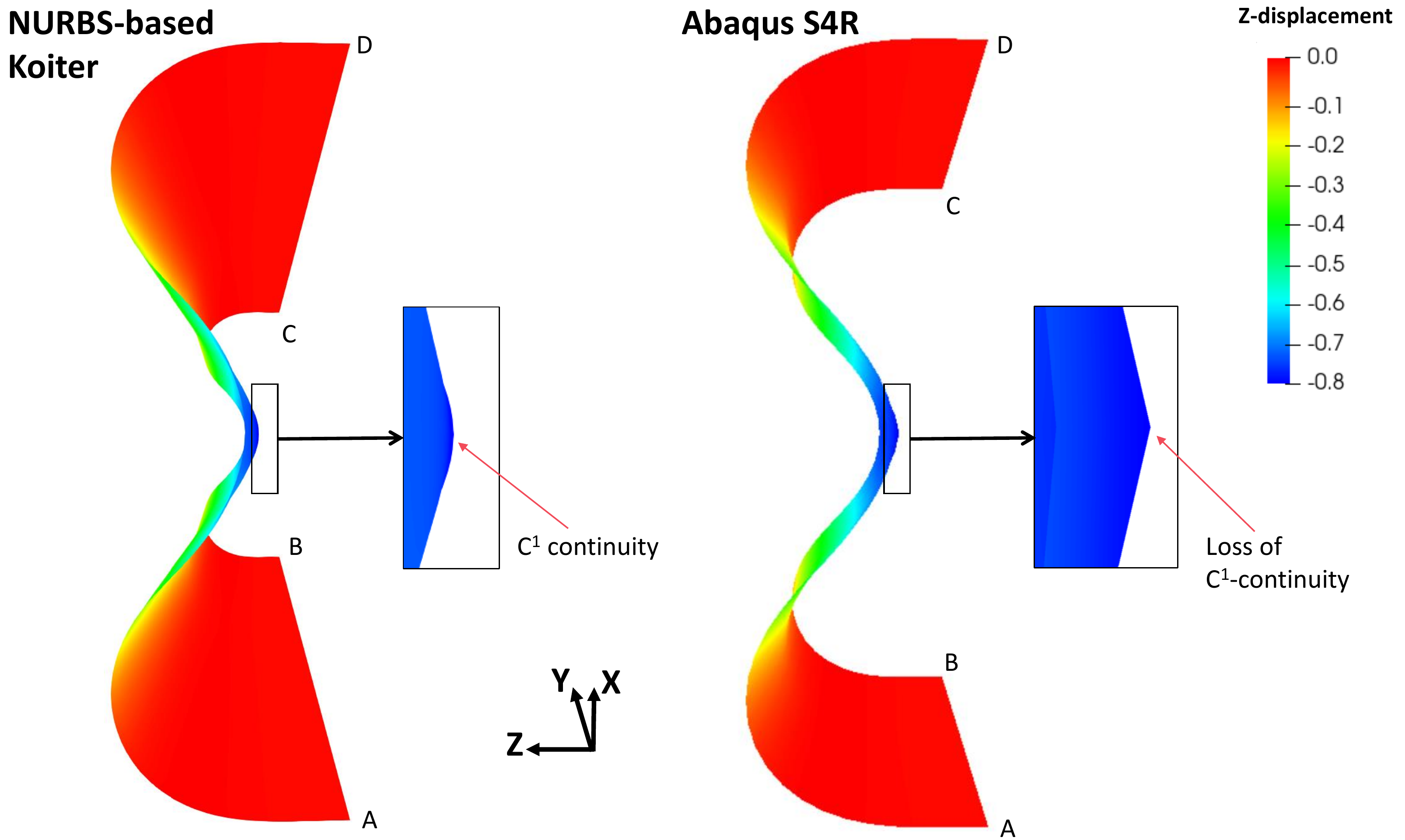}
\caption{Comparison of final equilibrium solutions between NURBS-based Koiter shell, and Abaqus S4R, under point indentation.}
\label{MSC_zoomed_midpoint}
\end{figure}


\clearpage
\subsection{Uniaxial tension of a rectangular sheet}
\begin{figure}[h!]
\centering
\includegraphics[scale=0.5]{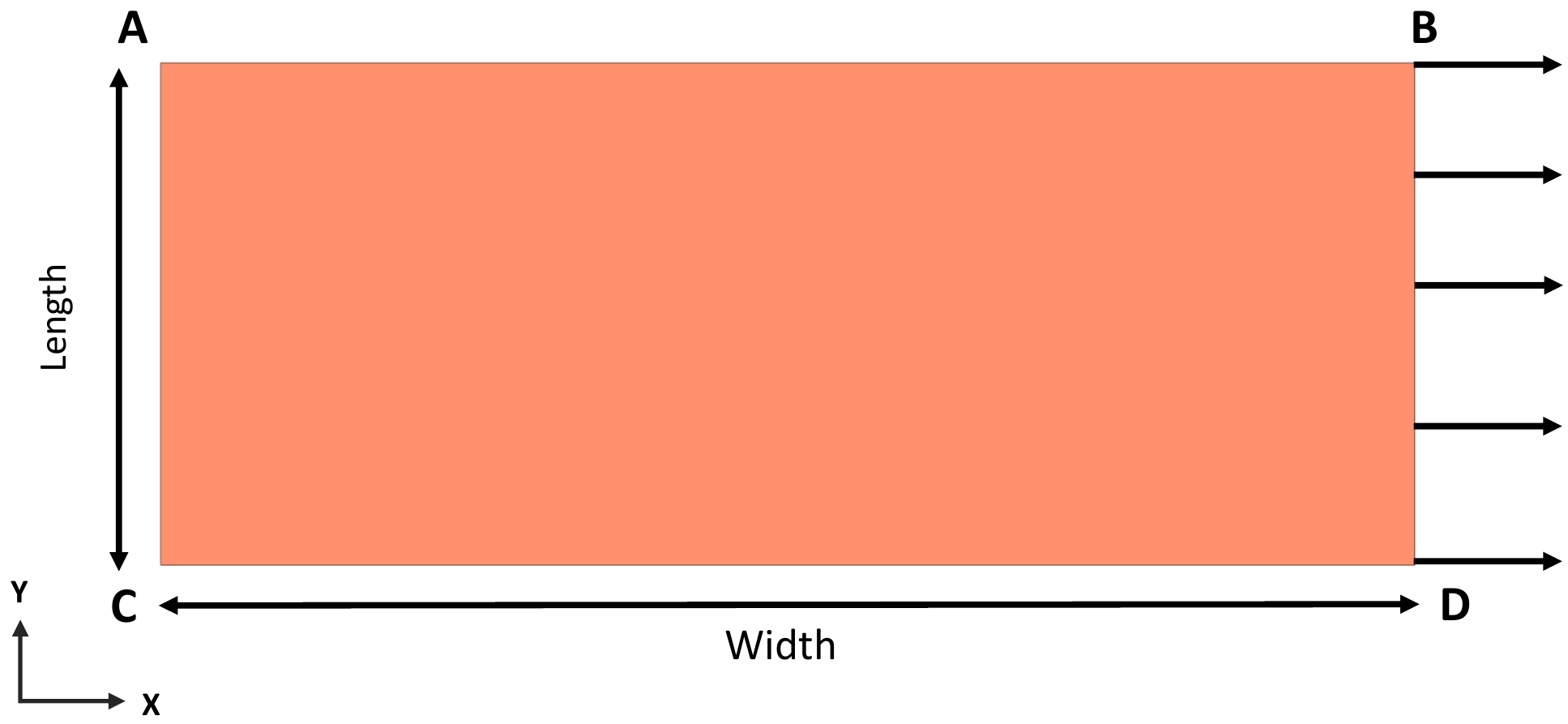} 
\caption{The reference configuration of a rectangular sheet under uniaxial tension.}
\label{Wrinkle_reference}
\end{figure}

The study of wrinkling in thin elastic sheets has attracted significant interest in the past two decades 
\cite{cerda2002wrinkling,cerda2003geometry,verhelst2021stretch,steigmann2013koiter,zheng2009wrinkling,healey2013wrinkling,puntel2011wrinkling,panaitescu2019birth},
and investigators have used a variety of two-dimensional models, numerical simulations, and even analytical treatments, to capture this 
phenomena. The most interesting feature of stretched-induced wrinkling of a thin sheet is the competition between bending and stretching that 
gives rise to wrinkles beyond a critical stretch, and then disappearance of wrinkles upon further stretching. Following \cite{cerda2002wrinkling,steigmann2013koiter,zheng2009wrinkling}, we study the wrinkling behavior of 
silicone rubber sheet with Young's modulus of 1.0 tonne mm$^{-1}$ sec$^{-2}$, and Poisson's ratio of 0.5 (numerically approximated as 
0.499999). Figure \ref{Wrinkle_reference} shows the reference configuration of a rectangular sheet which is subject to uniaxial tension,
in which CD is 254 mm, and AC is 101.6 mm. The sheet thickness is taken to be 0.1 mm. The edges AC and BD are constrained in the Y and Z 
directions, and the edge BD is displaced in the positive X direction by different amounts, that correspond to different levels of imposed nominal 
strains (7\%, 10\%, 15\%, and 20\% considered in this study). Pinned boundary conditions are applied at the edges AC and BD, and edges AB 
and CD are not allowed any Z-displacement. 

The rectangular sheet is discretized with 512 NURBS elements (16 along the length and 32 along the width), of degree 3, with 665 control points, 
each having unit weights. The DR parameters $m$, $c$, and $\Delta t$ are chosen as $0.000643$, $0.0001$, and $0.0001$, respectively.
The X-displacements on the edge BD are applied in 20 steps, where the intermediate steps are run for a fixed number of 2,500 iterations, and the last loading step is carried out with a termination criterion $\epsilon_t= 10^{-12}$. Random out-of-plane initial perturbations, in the range 
$[-0.1, 0.1]$, were given to the  sheet to trigger instabilities.  
\begin{figure}[h!]
\centering
\includegraphics[scale=0.5]{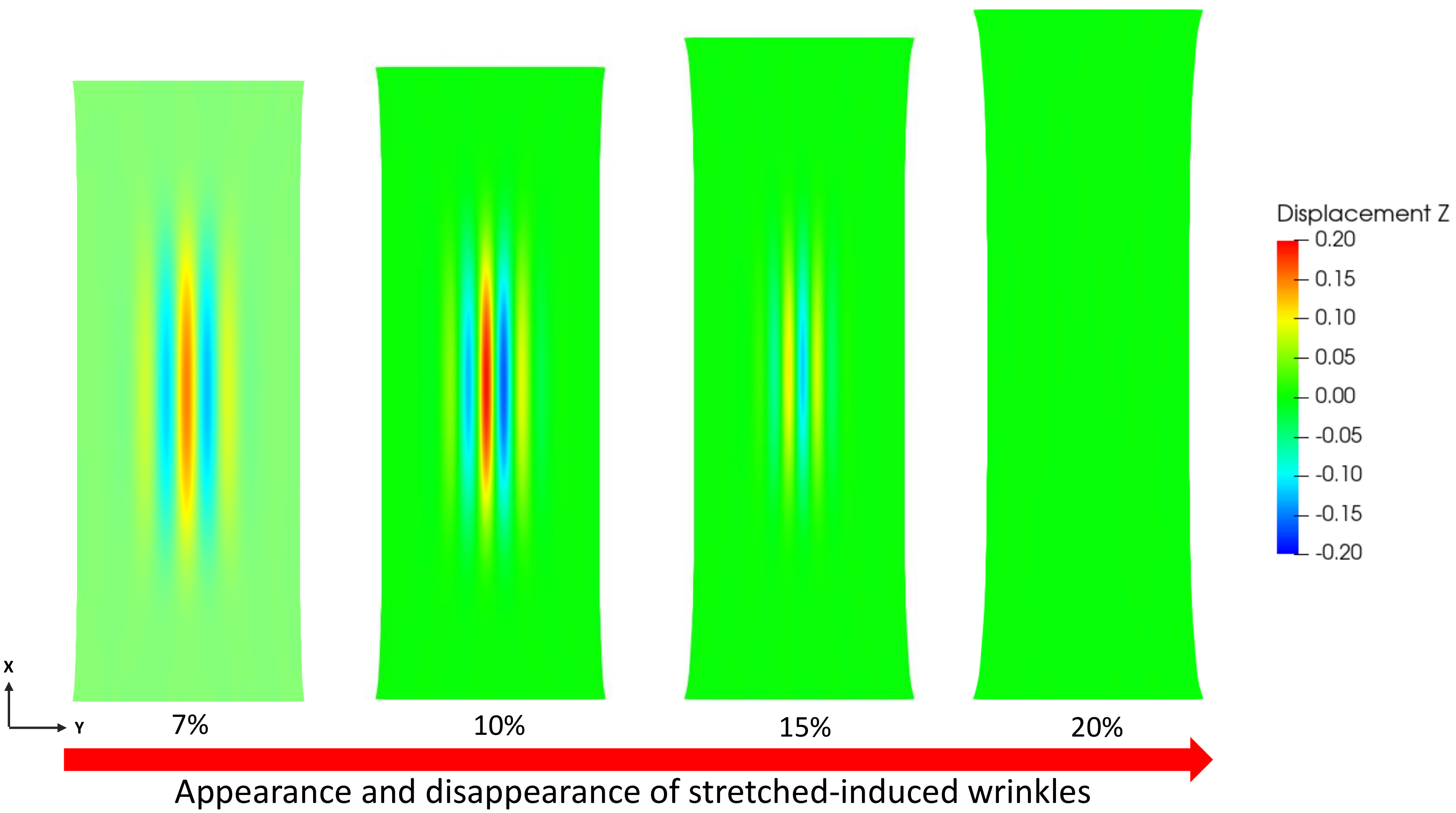} 
\caption{Wrinkles at different levels of applied nominal strains. Stretching first induces wrinkling, and then wrinkles disappear.}
\label{wrinkle_different_strains}
\end{figure}

\begin{figure}[h!]
\centering
\includegraphics[scale=0.35]{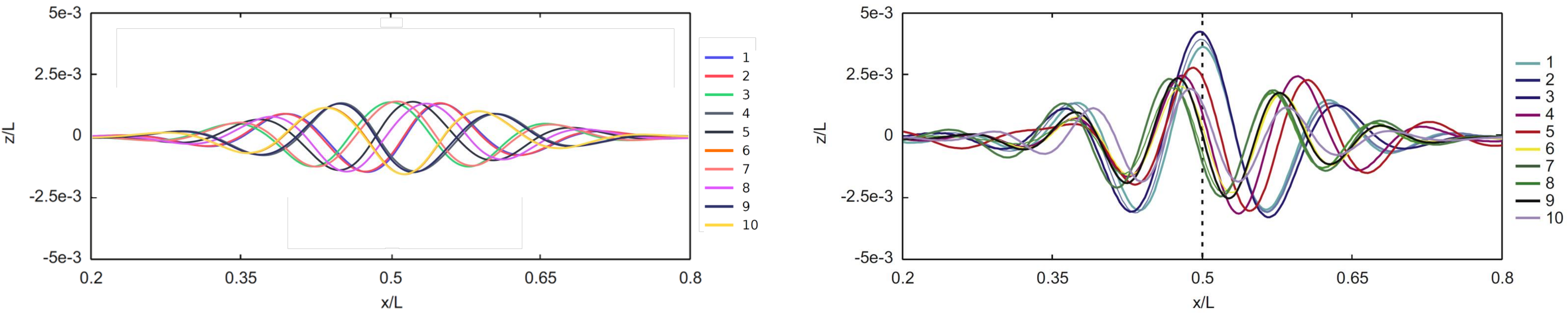} 
\caption{Different cross-sectional wrinkle profiles obtained with different random starting points for 10\% nominal strain (A) NURBS-based Koiter (B) Solutions reported by \cite{Taylor_Bertoldi_2014}}
\label{wrinkle_different_runs_10_perc}
\end{figure}

\begin{figure}[h!]
\centering
\includegraphics[scale=0.5]{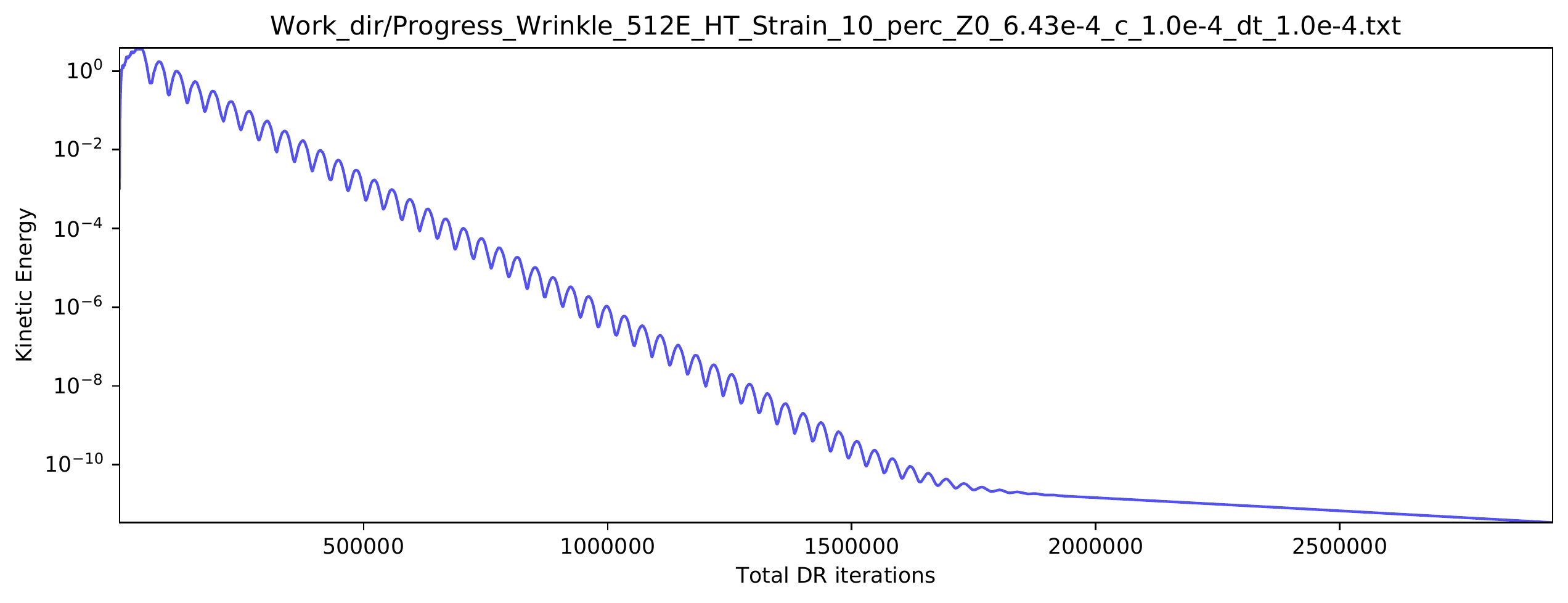} 
\caption{The evolution of kinetic energy during DR iterations for uniaxial tension.}
\label{wrinkle_ke_progress}
\end{figure}

Figure ~\ref{wrinkle_different_strains} shows growth of stretched-induced wrinkles in the thin sheet. The first appearance 
of wrinkles is noted at nominal strains past 5\%, the amplitude of wrinkles increases upon further stretching, and ultimately 
at large stretches the wrinkles disappear. These trends are consistent with those reported in the literature 
\cite{steigmann2013koiter,zheng2009wrinkling}. We note that the critical strain for the onset of wrinkling was identified 
as 3.8\% in \cite{zheng2009wrinkling}, where S4R Abaqus shell elements (employing finite membrane strains) were used. 
Sample evolution of kinetic energy with DR iterations for the 10\% nominal strain case is shown in Figure ~\ref{wrinkle_ke_progress}.
In \cite{steigmann2013koiter,healey2013wrinkling} a family of stable wrinkled states ranging from symmetric to anti-symmetric modes 
were reported to exist. We selected $10$ different random initial perturbations, and studied the formation of wrinkles at 10\% nominal strain,
and were able to obtain various wrinkled states. As discussed earlier, a nonlinear elasticity problem, in general, can exhibit multiple stable equilibrium states, and these different wrinkled states correspond to different equilibrium solutions. The cross-sectional profiles, at X = 127 mm, for different wrinkled states obtained from our formulation, and those obtained in \cite{steigmann2013koiter} are shown in Figure ~\ref{wrinkle_different_runs_10_perc}. Here, both the axes are non-dimensionalized 
by the length (101.6 mm).

Our trends are similar to those reported in the past, however, the peak height of the wrinkles is smaller in our case. A direct comparison of Z-displacement contour of our obtained symmetric mode, and that from 
\cite{steigmann2013koiter} is also shown in Figure ~\ref{wrinkle_different_runs_10_perc}. 
The reason for this difference can be attributed to the fact that in the classical Koiter nonlinear model,
a higher order term ($\zeta^2$) was dropped (Section ~\ref{shell-kinematics-derivation}, Equation ~\ref{Eab}) in estimation of the membrane 
strain. On the other hand, in \cite{steigmann2013koiter} authors have purportedly chosen a nonlinear Koiter model, however, they have not made 
any kinematic approximation in membrane strains as those made in Koiter's original nonlinear model. Similarly, Abaqus S4R element uses the full 
nonlinear membrane strain, and this fact could also 
explain the difference in our noted critical strain for the onset of wrinkling.

\begin{figure}[h!]
\centering
\includegraphics[scale=0.5]{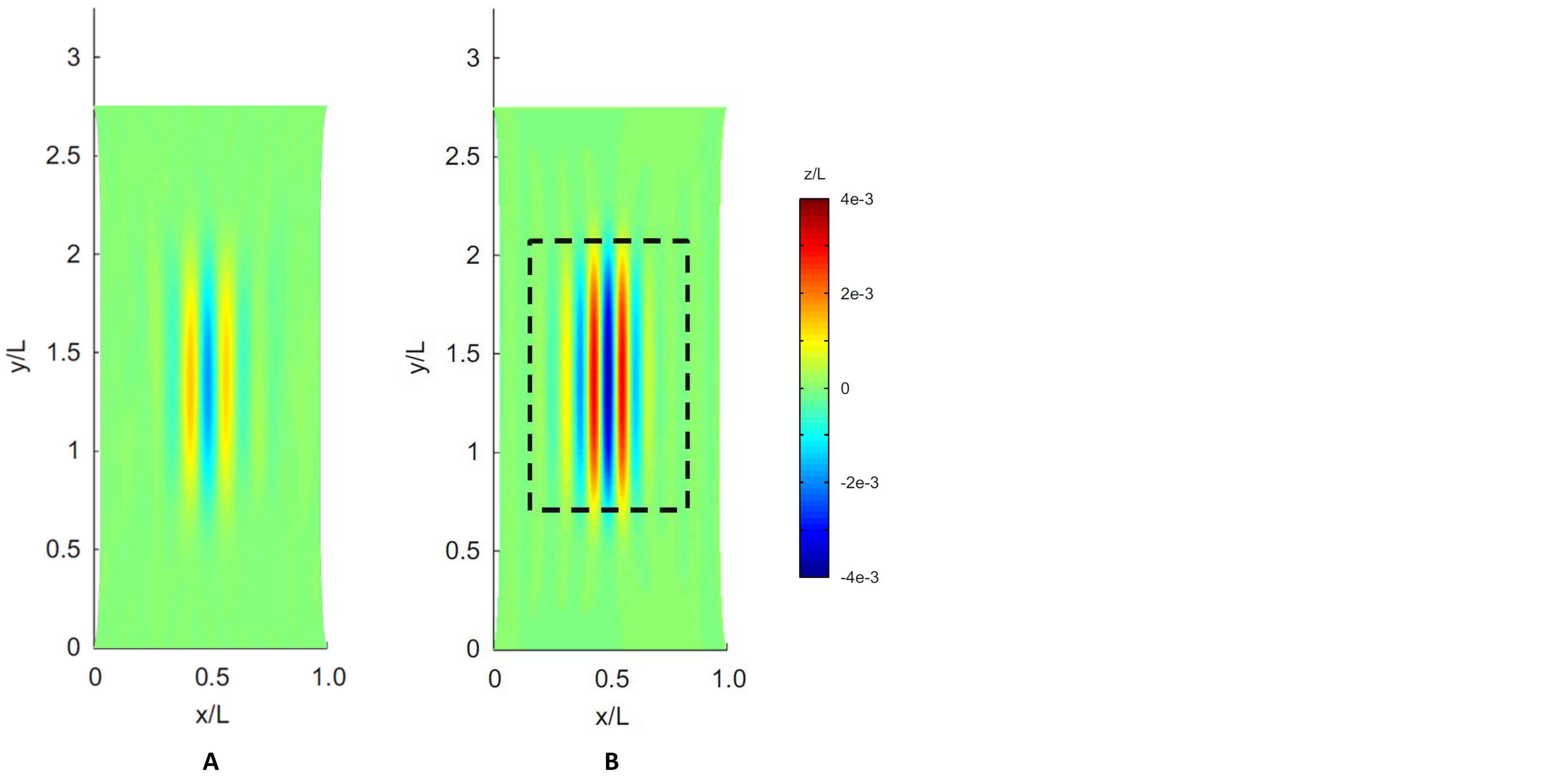} 
\caption{Comparing the Z-displacement contours for 10\% nominal strain between (A) NURBS-based Koiter (B) Solution reported by \cite{Taylor_Bertoldi_2014}.}
\label{wrinkle_compare_10_perc}
\end{figure}

\clearpage

\section{Conclusions}

Isogeometric analysis has been emerging as a competitive finite element technology, especially, since it is able to model 2D or 3D solid bodies 
exactly, in the sense of the CAD, and provides arbitrarily smooth basis that is highly effective in solving higher-order partial differential equations.  
In this work, we have developed a NURBS-based formulation for nonlinear Koiter shell to predict mechanical instabilities in problems involving wrinkling 
and buckling, adopted the method dynamic relaxation (DR) to solve for quasi-static equilibrium configurations, and developed a 
high-performance computing-based code architecture. Although the choice of explicit time integration scheme, used in DR, is limited by the numerical stability 
limits, it offers opportunity for exploiting the massive parallelization  offered by the modern computing architectures to enable speedy computations. 
Our proposed formulation exploits the Lyapunov stability of a mechanical system, i.e. relies on the principle that a dissipative mechanical system can converge to a stable 
equilibria once it is released from an initial state, and overcomes the traditional limitations of seeding imperfections in the initial 
structure (in an ad hoc manner) to trigger instabilities, or rely on computation of tangent stiffness matrix, which can become singular at the bifurcation 
point, for Newton-Raphson type iterative procedures. We noted that classical nonlinear Koiter model is able to effectively solve problems involving 
combined bending and stretching, and can generate highly complex wrinkling or buckling profiles in different scenarios. Although there is 
nothing sacrosanct about the Koiter nonlinear model per se, and it does not even have a general result for an existence of solution in the general nonlinear 
setting, it effectively captures, both, combined bending and stretching behaviors in thin shells, which is essential for capturing the phenomenon of wrinkling and buckling in 
several problems. 
We note that different modeling approximations can yield a trade-off between computational accuracy and efficiency, and in general, the most appropriate model can be problem-dependent; however, foregoing computational efficiency an ``all-round best'' nonlinear elastic shell model would still be desirable. These issues are being addressed in an ongoing work \cite{padhye-kalia-steigmann-model}. 
Other hyperelastic, and anisotropic, material models, or nonlinear shells can also be implemented in our proposed framework. The future 
applications of this framework would include elasto-plasticity, phase-field fracture, and pre-strained effects in shells.

\section{Acknowledgments}
Nikhil Padhye thanks John Hutchinson for providing the requested paper by Koiter during the Covid pandemic, and John Ball for helpful discussions.
David Parks is thanked for getting the first author interested in mechanics of thin-structures, and several fruitful discussions. 
\clearpage

\bibliographystyle{abbrv}
\bibliography{references}

\end{document}